\documentclass[12pt]{article}

\usepackage{scicite}
\usepackage{graphicx}
\usepackage{times}
\usepackage{amsfonts}
\usepackage{amsmath}

\newenvironment{sciabstract}{%
\begin{quote} \bf}
{\end{quote}}

\title{Programmable wave-based analog computing machine: a metastructure that designs metastructures} 

\author
{Dimitrios C. Tzarouchis,$^{1\dagger \ddagger}$ Brian Edwards,$^{1\dagger}$ Nader Engheta$^{1\ast}$\\
\\
\normalsize{$^{1}$Department of Electrical and Systems Engineering,}\\
\normalsize{School of Engineering and Applied Sciences,}\\
\normalsize{University of Pennsylvania, Philadelphia, 19104, U.S.A.}
\\
\normalsize{$^\dagger$These authors contributed equally to this work.}
\\
\normalsize{$^{\ddagger}$Present address: Meta Materials Inc. (Europe),}
\\
\normalsize{Ap. Pavlou 10A, Marousi, 15123, Greece.}
\\
\normalsize{$^\ast$To whom correspondence should be addressed; e-mail:  engheta@seas.upenn.edu}
}

\date{\today}

\begin{document} 


\baselineskip24pt


\maketitle


\begin{sciabstract}
  Abstract: 
  The ability to perform mathematical computations using metastructures is an emergent paradigm that carries the potential of wave-based analog computing to the realm of near-speed-of-light, low-loss, compact devices. 
  We theoretically introduce and experimentally verify the concept of a reconfigurable metastructure that performs analog complex mathematical computations using electromagnetic waves.
  Reconfigurable, RF-based components endow our device with the ability to perform stationary and non-stationary iterative algorithms.
  After demonstrating matrix inversion (stationary problem), we use the machine to tackle two major non-stationary problems: root finding with Newton’s method and inverse design (constrained optimization) via the Lagrange multiplier method.
  The platform enables possible avenues for wave-based, analog computations for general linear algebraic problems and beyond in compact, ultrafast, and parallelized ways.  
\end{sciabstract}

\paragraph*{One-Sentence Summary:}
  A reconfigurable wave-based analog computing metastructure that can inverse-design a metastructure.



Calculators of various kinds have emerged by forging numerical algorithms with corresponding technological platforms. 
While the algorithms describe the mathematical paths on how solutions to problems can be found, the platforms are responsible for the transliteration of this abstract path into measurable quantities. 
The algorithms, the platforms, and their fusion define such systems' features and limitations.  
Following the ever-growing demand for ultrafast, compact, low/near-zero-power, and integrable cyber-physical devices for mathematical computations, it is organic that significant research efforts focus on making these numerical systems as optimal and efficient as possible. 

This quest led to the exploration and development of unconventional analog computing systems that exploit electromagnetic waves to deliver parallelized, ultrafast, compact, low-power computations\cite{Caulfield2010, Solli2015, Wetzstein2020, Zangeneh-Nejad2020}.
The two main categories in this domain involve systems that utilize free-space (scattering) elements (e.g. lenses\cite{Wetzstein2020}), and waveguides (e.g. photonic systems\cite{Feldmann2021, Tegin2021} and phased arrays\cite{Sun2013}). 
Sufficient free space propagation can act as dense matrix multiplication\cite{Goodman}.
Realized with traditional optics, this results in bulky devices\cite{Vellekoop2007,Wetzstein2020}, while metasurfaces can be more compact\cite{Shaltout2019, Imani2020, Cheben2018}.
However, in both there can be major bottlenecks regarding photonic and electronic integration.  
Waveguiding systems offer more mature solutions for integrable and reconfigurable devices, at the expense of much larger footprints compared to their free-space counterparts. 
In all cases, their main challenge is reconfigurability since its implementation requires some form of a-priori mathematical calculations.
For instance, metasurfaces/complex media\cite{Matthes2019,Venkatesh2022} requires optimization, and photonic meshes require operator decomposition \cite{Reck1994,Miller2013,Clements2016,Marpaung2019}. 

In terms of their mathematical abilities, the above examples demonstrate wave-based analog computing with functionalities such as integration/differentiation in space\cite{Pors2015, Zhu2017, Cordaro2019, Zhou2020} and time\cite{Zangeneh-Nejad2019}, matrix-vector multiplication\cite{Macho-Ortiz2021}, emulating equations through physical phenomena\cite{Hughes2019, Vadlamani2020}, or acting as platforms for neural network functionalities\cite{Lin2018, Wetzstein2020, Tegin2021, Ashtiani2022}. 
The intersection with the metamaterial paradigm delivered a series of remarkable analog computing devices with matrix multiplication\cite{Silva2014, Pors2015, Zangeneh-Nejad2020} and ultimately equation solving (matrix inversion) capabilities\cite{MohammadiEstakhri2019}. 
In most of the cases the matrix computations (especially the matrix inversion\cite{MohammadiEstakhri2019}) were performed through stationary algorithms\cite{S.1995}, such as the Jacobi method, where the matrix (operator/kernel) does not change with the iteration count.

The fundamental and far-reaching question we address here is whether a wave-based analog metastructure can be reconfigurable simply and intuitively, without needing a-priori calculations.
Most importantly, the resolution to this question endows one with the ability to implement stationary and non-stationary algorithms.
We propose a device based on an RF waveguide architecture with reconfigurable components.
Regarding stationary problems, we use this device to perform matrix inversion of a statistically large number of matrices.
As for non-stationary problems, we demonstrate both root finding using Newton's method and Inverse Design.
All three examples are only possible due to the reconfigurability of the device and hint at the possibility of deeper explorations into the realm of advanced numerical algebra methods.


A conceptual representation of the main idea is pictorially summarized in Fig. 1 ({\bf A}). 
The main property of our proof-of-concept system distinctively different from all previous metastructure approaches (i.e. \cite{MohammadiEstakhri2019}) is its reconfigurability; the metastructure has the ability to rapidly take on different matrices (operators or kernels) $K$. 
To facilitate this, we employ a wave-based direct complex matrix (DCM) architecture, which offers an intuitive and simple implementation of any desired matrix~\cite{Tzarouchis2022}. 
Using waves instead of currents and voltages, it is analogous to the crossbar architecture used in electronic analog computing systems\cite{Ielmini2018,Zidan2018,Sun2019}, and it can be seen as a generalized phased array feed\cite{Sun2013}. 
In this device, a collection of $n\times n$ tunable phase shifting and amplifying elements (which can also act as attenuating element) connect an input vector of $n$ complex amplitudes on an array of transmission lines to a similar output vector through combiners.
This architecture can be seen schematically in Fig. 1 ({\bf B}) and its corresponding experimental implementation in Fig. 1 ({\bf C}).

The key component of the metadevice is the \emph{multiplier module} (Fig. 1 ({\bf D})) so named because given an input signal characterized by its complex amplitude at 45MHz, $V_\mathrm{in}$, it will render a similar output $V_\mathrm{out} = z V_\mathrm{in}$, where $z$ is a complex multiplication factor.
This module consists of two basic components: (a) a voltage-controlled phase shifter with over 360 degrees of potential phase rotation, and (b) a voltage-controlled amplifier with $\approx$47dB of dynamic range (-30dB to +17dB).
Through the use of an embedded microcontroller unit (MCU) in each multiplier, each device can be controlled externally through a suitable communication network and a computer (see supplementary material).
While the design frequency is 45MHz, the module could be implemented at RF (GHz) and photonic (THz) platforms platforms, following the same principle of operation.
The experimental DCM implementation consists of 25 multipliers to yield $5\times5$ complex matrices. 
The ingress and egress stages~\cite{Tzarouchis2022} are implemented with five 1-to-5 power splitters (ingress stage) and five 5-to-1 signal combiners (egress stage). 
The multipliers are clustered into five groups, one for each matrix row. 
In Fig. 1({\bf B}) we depict the planar schematic of the DCM suitable for photonic implementation. 
However, for the RF implementation we stacked and routed the components vertically (Fig. 1 ({\bf C})), making the device compact for our particular wavelength and platform choice. 
Different stacking or integrated circuitry approaches can potentially be used to further reduce its overall footprint.

The metadevice can be operated in one of two configurations with dramatically different results.
When the DCM is set in an open-loop configuration (Fig 1 ({\bf E}) inset), it can be used for rapidly calculating parallelized matrix-vector multiplication. 
However, a closed-loop configuration can be created by connecting the outputs and the inputs with a feedback loop using properly designed couplers (Fig 1 ({\bf F}) inset).
When the DCM is in a closed-loop configuration, the metadevice can rapidly calculate parallelized matrix inversion (equation solving).
This is a unique feature of metadevices/metastructures\cite{MohammadiEstakhri2019,Tzarouchis2022} that incorporate feedback loops. 

First we investigate the stationary analysis capabilities of our metadevice.
For the assessment of the open and closed loop operation, we performed a series of randomized trials, one instance of which is presented in Fig. 1 ({\bf E}) and ({\bf F}). 
For each trial, a random passive matrix $A\in\mathbb{C}^{5\times5}$ was chosen and applied to the metadevice in both its configurations.
Each measurement was performed by exciting each input port in turn with all other inputs appropriately terminated and then observing the complex amplitudes on each of the output ports.
For the open-loop configuration, this corresponds to performing five matrix-vector multiplications, or as $A\cdot I$ where each column of $I$ is progressively applied (one column at a time) as separate vectors.
The closed-loop configuration was measured similarly, but in this case we are probing the steady-state of the metadevice which corresponds to $(I-K)^{-1}\cdot I=A^{-1}\cdot I$.
While this measurement technique fully characterizes both configurations, in practice dense complex vectors will be input and read to achieve parallelized results.

The estimated relative error $||A_\textrm{exact}-A_\textrm{meas}||^2/||A_\textrm{exact}||^2 $ for both cases (Fig. 1 ({\bf E}) and ({\bf F})) revealed an error about 0.001 and 0.005, respectively.
Despite the component imperfections, misalignments, measurement noise, and other stochastic errors, the measured results are in excellent agreement with the theoretical values.
The calibration procedure of the metadevice and the statistical analysis of the full trial set (100 values) are presented in the Supplementary Material.

A single multiplier module has a rise time of approximately 80 ns to achieve its desired complex value. 
This value is approximately $4T$ assuming one-period duration of $T=1/45$MHz $\approx$ 22.2 ns. 
In the open loop configuration, the total response requires approximately $5T$, including signal delays in connections and splitters. 
The duration of the closed-loop case is affected by the platform and the condition number of the inverted matrix~\cite{Tzarouchis2022}, but in principle is in the same order of magnitude. 
Possible photonic implementations may further reduce this time to the picosecond range\cite{Ashtiani2022} and below\cite{Guo2022}.  


We now apply the implemented metadevice to two characteristic non-stationary problems that highlight its mathematical abilities: 
(i) root finding of a system of five equations with five unknowns using Newton’s iterative technique and (ii) implementing an inverse-design problem using the Lagrangian multiplier formalism for constrained optimization. 
Both cases require that the kernel be reprogrammed in each iteration step. 
Note that our approach is not restricted to these two problems; instead, we choose these to highlight the potential of the introduced metadevice.

For the first case we construct a simple nonlinear toy problem and we apply Newton's algorithm\cite{Bertsekas} (Fig. 2 ({\bf A})) for finding one possible root. The vector problem statement reads
\begin{equation}
 {\bf f}({\bf z})=\left[f_1({\bf z}),f_2({\bf z}),f_3({\bf z}),f_4({\bf z}),f_5({\bf z})\right]^T = {\bf 0}   
\end{equation}
where ${\bf f}\in\mathbb{C}^{5\times1}$, ${\bf z}=\left[z_1,z_2,z_3,z_4,z_5\right]^T\in\mathbb{C}^{5\times1}$, and $\bf 0$ is the zero vector.
We construct the vector function to have the following polynomial form 
\begin{align} 
    f_1({\bf z}) &= (z_1-r_1)(z_2-4.2i)(z_3+2)(z_4-5i)(z_5-3.5) \\ 
    f_2({\bf z}) &= (z_1-3.9)(z_2-r_2)(z_3+2.5i)(z_4-3.2i)(z_5-4.2) \\
    f_3({\bf z}) &= (z_1+5.2i)(z_2-4)(z_3-r_3)(z_4-4i)(z_5-7.1) \\
    f_4({\bf z}) &= (z_1-3)(z_2-7i)(z_3+4)(z_4-r_4)(z_5-5i) \\
    f_5({\bf z}) &= (z_1-5.2i)(z_2-4)(z_3+4.75i)(z_4-8)(z_5-r_5)
\end{align}
where ${\bf r} =\left[r_1,r_2,r_3,r_4,r_5\right]^T$  are the vertices of a regular pentagon with $1/4$ radius (see Fig. 2({\bf B})) and the other factors represent additional extraneous roots far from the starting point.

For the evaluation of Newton's method we need to calculate the Jacobian matrix, i.e., $J_{ij}=\frac{\partial f_i}{\partial z_j}$ or
\begin{equation}
{J}_f({\bf z}) = 
\begin{pmatrix}
\frac{\partial f_1}{\partial z_1} & \frac{\partial f_1}{\partial z_2} & \cdots & \frac{\partial f_1}{\partial z_5} \\
\frac{\partial f_2}{\partial z_1} & \frac{\partial f_2}{\partial z_2} & \cdots & \frac{\partial f_2}{\partial z_5} \\
\vdots  & \vdots  & \ddots & \vdots  \\
\frac{\partial f_5}{\partial z_1} & \frac{\partial f_5}{\partial z_2} & \cdots & \frac{\partial f_5}{\partial z_5} 
\end{pmatrix}
\end{equation}
therefore the root can be estimated by the following iterative process  
\begin{equation}
    {\bf z}_{n+1}={\bf z}_n-\alpha J^{-1}_f({\bf z}_n){\bf f}({\bf z}_n)
\end{equation}
where $\alpha=0.2$ is a relaxation constant~\cite{Tzarouchis2022}. 

In Fig. 2 ({\bf A}), we can see the required algorithm steps that implement the iterative scheme described by Eq. (8). 
Note that the Jacobian changes value in each iteration and it is required that its inverse is calculated anew.  
This is traditionally a computationally expensive operation which is accelerated through the use of our metadevice.
The results are then used to update the $\bf z$. 
The method converges successfully after a few iterations. 

A numerical version (using MATLAB) is compared with the experimental results illustrated in Fig. 2 ({\bf B}). 
We observe that for both MATLAB and the experiment, the estimation vector converges close to the exact roots.
Moreover, the estimated vector reaches a stationary point as the iteration count increases. 
After 15 iterations the relative error is $||{\bf z}-{\bf r}||^2/ ||{\bf r}||^2 \approx0.0023$.
This is similar to the accuracy achieved for the stationary trials, thus representing the accuracy floor of our system.
A similar picture is also visible by comparing three specific iterations, as illustrated in Fig. 2 ({\bf C}), where a comparison of the full Jacobian is presented.

The experimental results do not precisely follow the paths indicated by the numerical implementation realized using MATLAB. 
This can be explained by adding random noise to the Jacobian on each iteration step.
The added noise $N\in\mathbb{C}^{5\times5}$ is a random complex matrix that follows a normal distribution inside a disk with radius $r_N=0.01\lambda_\text{max}$, where $\lambda_\text{max}$ is the maximum eigenvalue of the Jacobian.
The noise creates many possible paths, all of which successfully converge and we observe that our measured results comfortably lie within these families of curves.
Note that some solution branches are more susceptible to this noise than others (e.g. $r_1$ (blue) and $r_3$ (red) curves in Fig. 2({\bf B}))  and this is due to the details of the toy problem solved.

Generally, the numerical accuracy of the device has a threshold that depends on both the implementation and the measuring apparatus (vector network analyzer (VNA)). 
When higher precision computations are required, this device can be a part of a mixed-precision computing system. 
In these systems, part of the calculations are done in a fast, low-precision estimation stage and then fed and further refined at a higher precision stage, similar to the in-memory mixed-precision approaches in electronic platforms~\cite{LeGallo2018}.

For the second example, we chose the case of an inverse design problem (Fig. 3 ({\bf A})). 
We assume that our design consists of a collection of $m=5$ two-dimensional (2D) scatterers with circular cross section at fixed known locations $\mathbf{r} = [r_1, ..., r_5]$, each with an unknown bounded permittivity $\mathbf{\varepsilon}=[\varepsilon_1, ..., \varepsilon_5 ] \in \mathbb{C}^{5\times1}$.
The goal is to achieve a specific user-defined scattered field measured at a series of $n=4$ detection (objective) points, $\mathbf{o} = [o_1, ..., o_4]$.
Note that in our case we assume a collection of cylindrical circular scatterers (2D) excited with a monochromatic incident field of $\lambda_\text{w}$ wavelength. 
The $x$-propagating incident field ($k_x$) is a polarized in the z-direction (TE wave - $E_z$) with the $e^{j \omega t}$ convention.

The scatterers are coupled, making this a nonlinear problem modeled using the Lippmann--Schwinger\cite{Pham2020} scheme, solved with a standard discrete dipole approximation (DDA) methodology~\cite{Yurkin2007}.
Each scatterer will respond to the local (self-excluded) electric field, which consists of the known incident electric field, $e_\text{inc}\in\ \mathbb{C}^{5\times1}$, and the scattered field from all other scatterers, $e_\text{sca}\in\ \mathbb{C}^{5\times1}$.
The scatterers exhibits a complex polarization vector $p = A(e_\mathrm{inc} + e_\mathrm{sca}) \in\ \mathbb{C}^{5\times1}$ where $A$ is the normalized polarizabilitiy diagonal matrix, i.e., $A=diag(\mathbf{\varepsilon}-\varepsilon_\text{background})$.
The field interaction between the scatterers are expressed via the Greens matrix $G\in\ \mathbb{C}^{5\times 5}$ (hollow symmetric matrix) such that $e_\mathrm{sca} = G p$.
We may express the polarization vector $p = A (e_\mathrm{inc} + G p)$, which indicates the mutual dependence of $p$.
Therefore, the polarization vector can be calculated as $p=(A^{-1}-G)^{-1} e_\text{inc}$.
Finally, we use the four objective points $\mathbf{o}$ to measure the scattered field vector $e_\text{meas} = G_\text{pr} p \in\ \mathbb{C}^{4\times1}$ where $G_\text{pr}\in\mathbb{C}^{4\times 5}$ is the propagator Greens function. 
The measured field is then compared to a (user-defined) objective $e_\text{obj}\in\ \mathbb{C}^{4\times1}$.

A typical constrained minimization problem (primal) can be written as~\cite{Bertsekas} 
\begin{equation}
    \begin{aligned}
        \min_{\substack{ x,y }} \quad & f(x,y)\\
        \textrm{s.t.} \quad  & g(x,y) \leq 0 
    \end{aligned}
\end{equation}
where $f(x,y)$ are the objectives and $g(x,y)$ are the constraints.
For such problems the Lagrangian (dual) problem is expressed as 
\begin{equation}
    \begin{aligned}
        \max_{\lambda}\min_{\substack{x,y}} \quad & \mathcal{L}(x,y,\lambda) = f(x,y) + \lambda g(x,y)
    \end{aligned}
\end{equation}
Note that $x$ and $y$ may be subject to further requirements such as domains and bounds.

For our particular example we have that $x=p$, $y=\varepsilon$, and $f(p,\varepsilon) = 1/2||G_\text{pr}p-e_\text{obj}||^2$ and $g(p,\varepsilon) = 1/2||(A(\varepsilon)^{-1}-G)p-e_\text{inc}||^2$.
In our formulation, the objective is the scattered field at the observation points. 
The constraints comprise the self-consistency of the polarization vector (physics). 
Also, the permittivity vector is subject to specific bounds, i.e., $\varepsilon \in \mathbb{R}$ and $\varepsilon \in [1,5]$.
Note that $g(p, \varepsilon)$ is nonlinear with respect to $p$ and $\varepsilon$ and therefore requires a non-stationary approach.

Following an initialization, our numerical evaluation of the above is implemented by a non-stationary algorithm that requires repeated application of the following three steps.
First, we minimize with respect to $\varepsilon$ by examining $\nabla_\varepsilon\mathcal{L}(p,\varepsilon,\lambda)=0$. 
At this step we project the resulting permittivity vector to the desired domain and bounds.
Second, we minimize with respect to $p$ by examining $\nabla_p\mathcal{L}(p,\varepsilon,\lambda)=0$. 
At this stage the required stationary matrix inversion is performed with our metadevice.
Finally we maximize for $\lambda$ by using $\nabla_\lambda\mathcal{L}(p,\varepsilon,\lambda)=0$.
These steps are repeated until convergence is achieved, i.e., 
\begin{equation}
\mathcal{E} = ||e_\text{obj}-e_\mathrm{meas}||^2/||e_\text{obj}||^2<\delta    
\end{equation}
(for more information see SM).

As a numerical test case, the scatterers are assumed to be lossless with permittivity of $\varepsilon=\left[\varepsilon_1,\varepsilon_2,\varepsilon_3,\varepsilon_4,\varepsilon_5\right]=[3.5,\ 1.5,\ 1.5,\ 3.5,\ 1.5]$.
The objective scattered field at the detection points $\mathbf{o}$, as depicted in (Fig. 3({\bf A})), is $e_\mathrm{obj} = [-0.0086 - 0.0078j, 0.0089 - 0.0132j, -0.0066 - 0.0120j, 0.0043 - 0.0004j]$. 
The values were extracted from the DDA method and verified with a full-wave COMSOL simulation. 
Note that the Fig 3({\bf A}) depicts the complex (hue/saturation) of the electric scattered field ($E_z$), i.e. the difference between the total field and the incident excitation.  

Figure 3 ({\bf B}) depicts a set of four cases for the same algorithm. 
In the first case (black line), the idealized (noiseless, no filtering) computer evaluation of the algorithm is given - we observe that after only 20 iterations the error drops below $10^{-3}$.
The experimental results are presented in Fig. 3({\bf B}) as red dots.
The measured results exhibit an optimal point (minimum error) after 87 iterations, with an error of $0.00172$. 
As an analog device, there is an additional systematic/stochastic/experimental noise to the system which affects the fidelity of the matrix inversion.
We apply a simple averaging filtering scheme on the polarization estimation, i.e., $p_\text{new}=(1-\alpha_\text{F})p + \alpha_\text{F}p_\text{previous}$, with $\alpha_\text{F}=0.25$, as a way to partially mitigate this noise. 
The filter affects the convergence speed by increasing the iteration count but also significantly improves the accuracy/fidelity of the matrix inversion, hence the metadevice's performance.
This feature is illustrated in Fig. 3 ({\bf B}), where the retrieved experimental results are compared to the idealized computer evaluation with the applied filter (blue line).
We also performed a series of 100 randomized cases of the idealized filtered computer evaluation with added noise to the estimated/measured polarization vector (faint blue lines in Fig. 3(B)).
The noise profile is similar to the one used in the first example (Newton's method).
The measured results are well contained within these error bounds.
Note that iteration count is not equivalent of time.
For a traditional computer evaluation, each iteration (with its required matrix-inversion) could ultimately be slower than the convergence time of an optimized hardware implementation of the metadevice.

Due to systematic/measurement noise, the error begins to grow after the experimental accuracy floor is obtained - an indication that a termination criterion could be applied at this point. 
This result also agrees with the maximum accuracy we obtained in the previous non-stationary example.
More sophisticated error-correcting and filtering schemes can possibly push the accuracy below this threshold.
For instance, $\alpha_\text{F}$ could be adaptively tuned during the non-stationary evaluation to realize a mixed-precision computing system.

At the minimum error point (iteration 87), the extracted permittivity estimation is illustrated in (Fig. 3({\bf C})). 
Notice that the values are very close to the numerical test case objectives and permittivities. 
Finally, Fig. 3 ({\bf D}) illustrates the path of the scattering vector, $e_\text{meas}$, for these 87 iterations. 
Similar to the above example, the faint paths represent the added noise effects to the numerical evaluation. 

For both presented non-stationary examples, it is evident that our metadevice can act either as an ultrafast analog computing machine and mathematics calculator with waves, or in a broader sense as an electromagnetic emulator for inverse design\cite{Molesky2018}. 
It can be used for a plethora of realistic problems where the linear response of a system (i.e. matrix-vector multiplication) or the solution of a system of equations (stationary problems, matrix inversion) is required.  Moreover, the intuitive reconfigurability of this metadevice also enables the performance of constrained optimization tasks, like the ones required in non-stationary problems such as inverse design, where the desired response of complex media requires intensive optimization~\cite{Horodynski2022}. In short, this metastructure can design metastructures.
Finally, an adaptation of the above proof-of-concept metadevice in RF-IC, photonic, or hybrid platforms can make it an excellent candidate for on-the-fly or computation-through-propagation ultrafast, parallelized calculations.  

\bibliography{scibib}
\bibliographystyle{Science}

\section*{Acknowledgments}
The authors would like to thank Mario Junior Mencagli for useful discussions and preliminary experimental survey on the subject. D.C.T acknowledges Luiz F. O. Chamon and Juan Cervi\~no for the useful inputs and discussions regarding the constrained optimization algorithm. 
\paragraph{Funding}: This work is supported in part by the Air Force Office of Scientific Research (AFOSR) Multidisciplinary University Research Initiative (MURI) grant numbers FA9550-17-1-0002 and FA9550-21-1-0312.
\paragraph{Competing Interests}: N.E. is a strategic scientific advisor/consultant to Meta Materials Inc.  The authors have no competing interest. 
\paragraph{Authors Contributions:} N.E. conceived the idea for the reconfigurable device that solve equations, acquired the funds, and supervised the project. D.C.T developed further the relevant theories and analyses of the project. B.E. designed and programmed the device and the device's calibration routine. D.C.T. and B.E. assembled, built, tested the components, and performed simulations and experimental measurements. D.C.T and B.E. developed the numerical examples. All the authors discussed the results.  D.C.T wrote the first draft of the manuscript and D.C.T, B.E. and N.E. discussed, developed, and edited the final version of the manuscript.
\paragraph{Data and materials availability:} All data needed to evaluate the conclusions in the paper are present in the main text and the supplementary materials.


\section*{Supplementary Materials}
Materials and Methods\\
Supplementary Text\\
Figs. S1 to S9\\
References \textit{(1-15)}

\begin{figure}
\centering
    \includegraphics[width=1\textwidth]{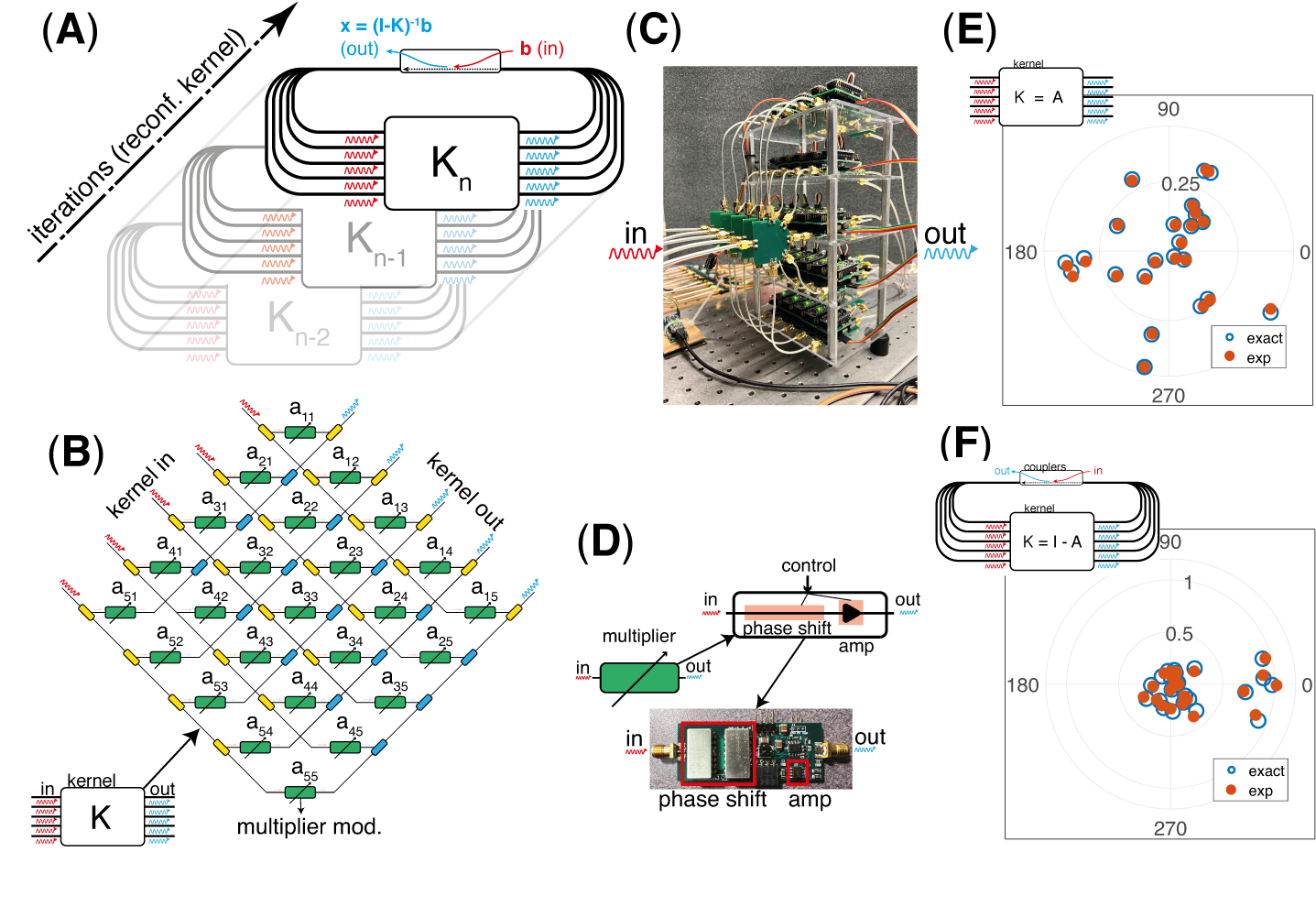}
    \caption{{\bf 
    A reconfigurable wave-based analog computing metastructure}: 
    (A) A conceptual representation that describes the main objective, i.e., a reconfigurable device that can provide us with repeated matrix inversions of arbitrary matrices in order to achieve stationary and non-stationary algorithms.
    The device can implement any given kernel $K=I-A$ and give the $\left(I-K\right)^{-1}=A^{-1}$. 
    (B) The central component of the design consists of a direct complex matrix (DCM)~\cite{Tzarouchis2022} architecture of 5x5 elements. 
    (C) The experimental realization of this design for the 45 MHz operating frequency. 
    (D) The essential element of the DCM is the multiplier module, consisting of both a phase-shifting and an amplification part (which can also function as attenuation part); controlled with an onboard microntroller unit. 
    Finally, (E) and (F) depict the performance of the DCM machine in the open-loop (matrix-vector multiplication) and the closed-loop (matrix inversion) setups.
    In both we compare the experimentally obtained matrix to one computed conventionally to see good agreement.
    }
    \label{fig:1}
\end{figure}

\begin{figure}
\centering
    \includegraphics[width=1\textwidth]{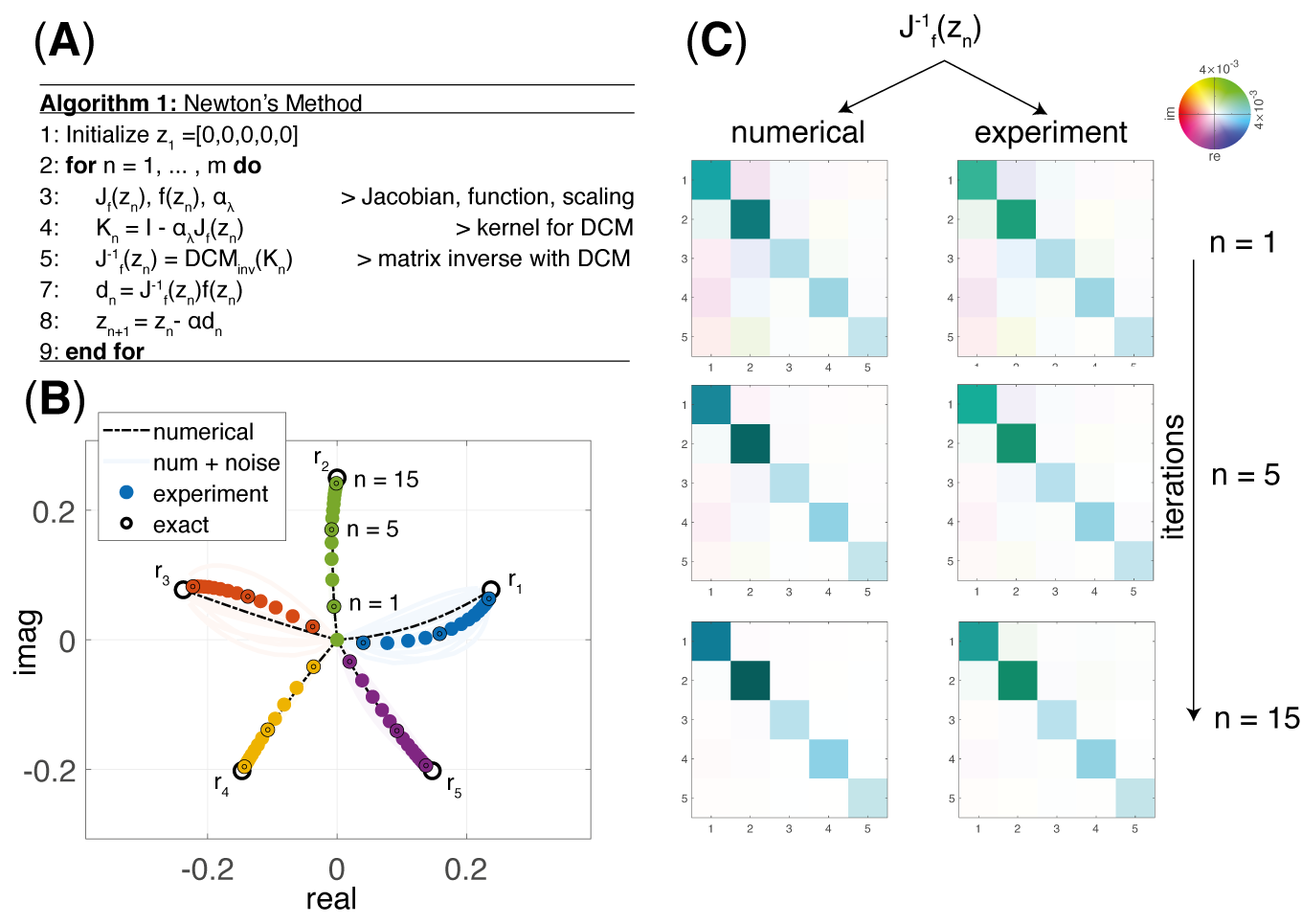}
    \caption{{\bf 
    Experimental verification of Newton’s root finding method with the proposed metadevice}: 
    (A) The algorithmic steps implemented in Newton’s root finding method. 
    A fixed kernel is programmed into the DCM machine in each iteration. 
    The measured results are used for calculating the next steps of the algorithm. 
    (B) A comparison between the experimental results and a numerical implementation of the algorithm. 
    The faint solid lines are cases where additional stochastic noise has been added to the system. 
    We observe that the experimental trajectory is well contained within these simulated noisy paths.
    (C) A comparison between the numerical (left column) and measured (right column) evaluation of the inverse Jacobian at different iterations.}
    \label{fig:2}
\end{figure}

\begin{figure}
\centering
    \includegraphics[width=1\textwidth]{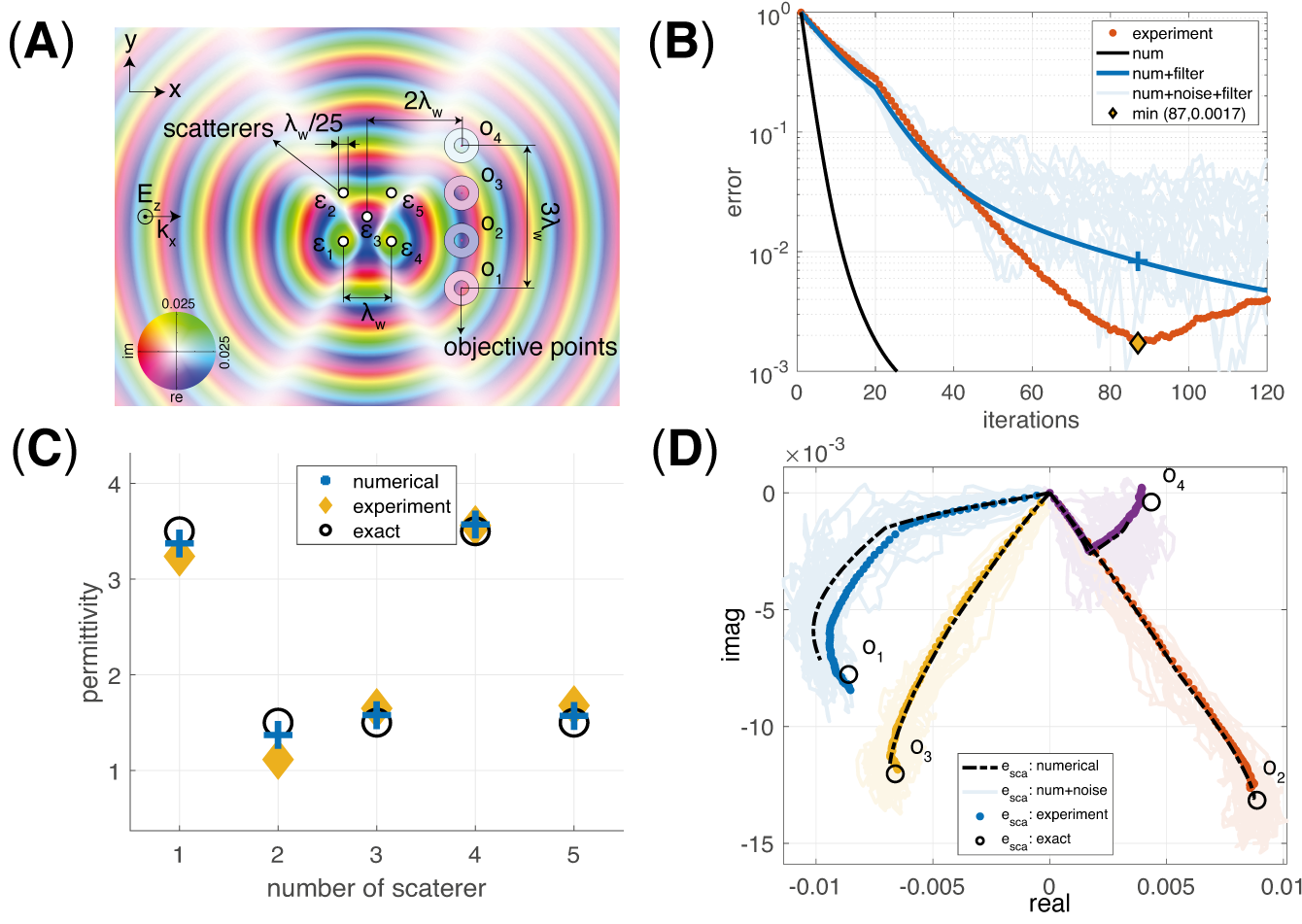}
    \caption{{\bf A metastructure that designs a metastructure: Numerical and experimental results}: 
    (A) Schematic of the numerical test case. A set of five two-dimensional (2D) scatterers with unknown permittivities, to be determined via our analog metadevice, $\varepsilon = \left[\varepsilon_1, ..., \varepsilon_5\right]$ subject to a plane wave excitation.
    Pictured is the complex-valued scattered field.
    The scattered fields at the observations points $[o_1,...,o_4]$ are used as the benchmark values of this problem in which the objective fields are shown as the color in each torus.
    The algorithm tunes each permittivity in order to match the scattered field (center of each torus) to the objective fields.
    (B) The relative error $\mathcal{E}$ for the algorithm computed both numerically, and experimentally using the metadevice, under various noise and filtering scenarios.
    The experimental device gives a minimum relative error of 0.00172 at 87 iterations.
    (C) Objective permittivities (black rings) compared to those computed numerically (blue cross) and experimentally via the metadevice (yellow rhombus) using the described algorithm at iteration 87. 
    (D) Evolution of scattered field vector up to iteration 87 in comparison to objective fields for experiment and simulation under various noise and filtering scenarios.
    }
    \label{fig:3}
\end{figure}


\clearpage


\end{document}


\maketitle


\section{Details for the root finding algorithm}
In the following the lowercase quantities are vectors, the capitalized ones are matrices while Greek letters denote scalars - the subscripts follow a logical notation. The problem statement for the root finding procedure is  
\begin{equation}
    f({\bf z})=0
\end{equation}
find ${\bf z}\in\mathbb{C}^{m\times1}$ that satisfies the above equation (roots) of $f\in\mathbb{C}^{m\times1}$. The above problem is solved using Newton's method for finding the root of a vector polynomial function~\cite{Bertsekas}. For example we have  
\begin{equation}
{f}({\bf z}) = 
\begin{pmatrix}
f_1(z_1,z_2,z_3,z_4,z_5) \\
f_2(z_1,z_2,z_3,z_4,z_5)  \\
f_3(z_1,z_2,z_3,z_4,z_5)  \\
f_4(z_1,z_2,z_3,z_4,z_5)  \\
f_5(z_1,z_2,z_3,z_4,z_5) 
\end{pmatrix}
\end{equation}
with $z_{1\cdot 5}\in \mathbb{C}$ or (equivalently)  
\begin{align} 
    f_1({\bf z}) &= (z_1-r_1)(z_2-4.2i)(z_3+2)(z_4-5i)(z_5-3.5) \\ 
    f_2({\bf z}) &= (z_1-3.9)(z_2-r_2)(z_3+2.5i)(z_4-3.2i)(z_5-4.2) \\
    f_3({\bf z}) &= (z_1+5.2i)(z_2-4)(z_3-r_3)(z_4-4i)(z_5-7.1) \\
    f_4({\bf z}) &= (z_1-3)(z_2-7i)(z_3+4)(z_4-r_4)(z_5-5i) \\
    f_5({\bf z}) &= (z_1-5.2i)(z_2-4)(z_3+4.75i)(z_4-8)(z_5-r_4)
\end{align}
where 
\begin{equation}
r=\frac{1}{4}\left(s_1+c_1i,-s_1+c_1i,-s_2-c_2i,s_2-c_2i,1i\right)^T
\end{equation}
with $c_1=\cos(2\pi/5)$, $c_2=\cos(\pi/5)$, $s_1=\sin(2\pi/5)$, and $s_2=\sin(4\pi/5)$. The point corresponds to the vertices of a regular pentagon. For the evaluation of Newton's method we need to calculate the Jacobian matrix, i.e., $J_{ij}=\frac{\partial f_i}{\partial x_j}$ (here $i$ and $j$ are indexes) or
\begin{equation}
{J}_f({\bf z}) = 
\begin{pmatrix}
\frac{\partial f_1}{\partial z_1} & \frac{\partial f_1}{\partial z_2} & \cdots & \frac{\partial f_1}{\partial z_5} \\
\frac{\partial f_2}{\partial z_1} & \frac{\partial f_2}{\partial z_2} & \cdots & \frac{\partial f_2}{\partial z_5} \\
\vdots  & \vdots  & \ddots & \vdots  \\
\frac{\partial f_5}{\partial z_1} & \frac{\partial f_5}{\partial z_2} & \cdots & \frac{\partial f_5}{\partial z_5} 
\end{pmatrix}
\end{equation}
therefore the root can be found as 
\begin{equation}
    {\bf z}_{n+1}={\bf z}_n-\alpha J^{-1}_f({\bf z}_n)f({\bf z}_n)
\end{equation}
where $\alpha$ is a relaxation constant. Here we used $\alpha = 0.2$. In terms of an algorithm, we have the following routine

\begin{algorithm}
\caption{Root finding with Newton's method}
\begin{algorithmic}[1]
    \State Initial guess for ${ z}_1$  
    \For {$n=1,\dots,m$}
    \State \textcolor{black}{${ J}_{f}({\bf z}_n)$}
    \State \textcolor{black}{$\alpha_\lambda=\frac{2}{\lambda_{min}+\lambda_{max}}$} \Comment{Scaling factor: $\lambda_{mim/max}$ are the min/max eigenvalues of ${ J}_{f}({\bf z}_n)$}
    \State \textcolor{black}{$K_n= I-\alpha_\lambda{ J}_{f}({\bf z}_n)$} \Comment{Kernel that is fed to DCM machine}
    \State ${\bf d}_n = {J}^{-1}_{f}({\bf z}_n){f}({\bf z}_n)$  \Comment{Compute matrix inverse with the DCM machine}
    \State ${\bf z}_{n+1}={\bf z}_{n}-\alpha{\bf d}_n$
    \EndFor
\end{algorithmic}
\end{algorithm}

\section{Details on the inverse design algorithm}
In this section we present the details for the inverse design algorithm implemented in text. 
The algorithm consist of a part of the DDA methodology for the quantification of the problem and an its adaptation to a Lagrange formalism for solving the require inverse scattering problem, the determination of the permittivity of the scatterers. 
Note that both methods are arguably the simplest methods to follow, since they offer an intuitive understanding on the formulated problem and the coorresponding inverse-design (constraints optimization) problem.

\subsection{Notes on the DDA method}
In this section we present a few details regarding the DDA method used in the main text. 
The details can be found also in~\cite{Purcell1973,Draine1994,Yurkin2007,Groth2020}. 
A similar methodological approach was also used in~\cite{Nikkhah2022}.

We start by assuming that each 2D scatterer (assuming a point in the x-y plane) acquires its z-oriented dipole moment due to the local electric field, i.e.,
\begin{equation}\label{Pol_Efield}
    p=\alpha e_\text{loc}
\end{equation}
where the $e_\text{loc}$ is the vector of local z-polarized electric fields at the center of each point and $\alpha$ is the polarizability that depends on the shape and the material composition of each 2D scatterer. 
The local field is the sum of the incident field and the secondary fields generated from all the other dipoles such that:  
\begin{equation}\label{localEfield_Pol}
    e_\text{loc}=e_\text{inc}+G p
\end{equation}
where $e_\text{inc}$ is the incident field vector, ${p}$ is the induced polarization vector and ${G}$ is the 2D Green's function. 
In our case we consider a two-dimensional (2D) problem with a transverse electric (TE) excitation (the field is normal (z-direction) to the x-y plane). 
Therefore the corresponding Green's function reads
\begin{equation}\label{hankelGreen}
    {G} = {G}({r}_i-{r}_k)=-j\frac{k_0^2}{4\pi \varepsilon_0}H^{(2)}_0(k_0|{r}_i-{r}_k|)
\end{equation}
where $H^{(2)}_0(k_0|{r}_i-{r}_k|)$ is the Hankel function of the second type (with the time harmonic convention $e^{+j\omega t}$) and 0-th order and $k_0 = \omega_0 \sqrt{\mu_0 \varepsilon_0}$ is the free-space wavenumber~\cite{balanis1999advanced}. 
The $G$ is a $\mathbb{C}^{N\times N}$ Toeplitz matrix with zero diagonal entries since the $|{r}_i-{r}_k|$ is treated as in (assuming a uniformly spaced discrete grid)~\cite{Purcell1973,Draine1994,Yurkin2007,Groth2020}. 

By combining Eqs.~(\ref{Pol_Efield}) and~(\ref{localEfield_Pol}) we obtain the following expression, arranged using the matrix formulation as follows
\begin{equation}\label{pol_Einc}
    p=\left(A^{-1}-G\right)^{-1}e_\text{inc}
\end{equation}
where the lowercase quantities $p=[{p}_1,{p}_2,...,{p}_N]^T$, $A=\text{diag}(\alpha)$, $\alpha=A_{\text{cell}}\varepsilon_0[\varepsilon_1-1,\varepsilon_2-1,...,\varepsilon_N-1]$ ($A_{\text{cell}}$ is the cross-sectional area of a cylinders) and $e_\text{inc}=[e^\text{inc}_1,e^\text{inc}_2,...,e^\text{inc}_N]^T$ are $\mathbb{C}^{N\times1}$ vectors, $\text{diag}(\cdot)$ is the diagonal matrix operator.

Finally, the scattered field observed at $M$ specified discrete detection points (in general $M\neq N$) is given by: 

\begin{equation}\label{Escat_Einc}
    e_\text{sca}=G_\text{pr}~p=G_\text{pr}\left(\text{diag}(\alpha^{-1})-G\right)^{-1}e_\text{inc}
\end{equation}
where $G_\text{pr}\in\mathbb{C}^{M\times N}$ is the ``propagator" Green's function matrix. This propagator function connects the induced dipole polarization vectors of the scatterers with the desired detection (or objective) points. The above matrix representation of the scattering problem allow us to have a clear inspection of the unknown quantities. These quantities are the ones that will be formulated as a Lagrange multiplier algorithm for the solution of the desired constrained optimization problem. 
We note here that, as seen in Eq.~(\ref{Escat_Einc}), the forward scattering problem requires a matrix inversion to evaluate the polarization density vectors induced in each scattering cell, as we have discussed in our previous work~\cite{Nikkhah2022} in which we utilized the same DDA approach for the evaluation Eq.~(\ref{pol_Einc}) and the matrix-vector operation of Eq.~(\ref{Escat_Einc}) for different excitation and for different scattering scenarios.

\subsection{Notes on the Lagrange multiplier algorithm}

For this, we utilize the DDA algorithm (where we closely follow the contrast source inversion method!\cite{Berg1997}) and the Lagrange multiplier method for applying the constraints and finding the optimal solution. First, in terms of the defined problem, we have that the polarization is connected with the following expressions 
\begin{equation}
    p = A(e_\text{inc}+Gp)   
\end{equation}
and 
\begin{equation}
    e_\text{sca} = G_\text{pr}p
\end{equation}
A typical constrained minimization problem (primal) can be written as~\cite{bertsekas2009convex,Bertsekas,boyd2004convex} 
\begin{equation}
    \begin{aligned}
        \min_{\substack{ x,y \\ y \in \mathbb{R} \\ 0 \leq y \leq 1 \\ }} \quad & f(x,y)\\
        \textrm{s.t.} \quad  & g(x,y) \leq 0 
    \end{aligned}
\end{equation}
where $f(x,y)$ is the objective and $g(x,y)$ are the constraints also subject to further requirements of the problem such as $ y \in \mathbb{R} $ and $0 \ge y \ge 1 $
For such problems the dual Lagrangian problem is expressed as 
\begin{equation}
    \begin{aligned}
        \max_{\lambda}\min_{\substack{x,y \\ y \in \mathbb{R} \\ 0 \leq y \leq 1 \\ }} \quad & \mathcal{L}(x,y,\lambda) = f(x,y) + \lambda g(x,y)
    \end{aligned}
\end{equation}
which is essentially a dual unconstrained problem (since all the constraints are encapsulated to the $\lambda$ term). It is worth noting that the Lagrange multiplier should be positive real, $\lambda \in R^+$.
Finding an approximate solution to the primal inverse scattering problem is therefore reduced to finding a solution to the above dual problem. 
Notice that the Lagrange multiplier can be applied to either $f(x,y)$ or $g(x,y)$ without affecting the outcome of the overall process. 

The algorithm for solving the above dual problem is the following:
\begin{itemize}
    \item Step 0: initial $x_0$ and $\lambda_0$
    \item Step 1: minimize $y_n$, i.e.,  via $\nabla_y\mathcal{L}(x_{n-1},y_{n},\lambda_{n-1}) = 0$
    \item Step 2: project $y_n$ into $y \in \mathbb{R}$ and $0 \leq y \leq 1$
    \item Step 3: minimize $x_n$, i.e., $\nabla_x\mathcal{L}(x,y_n,\lambda_{n-1}) = 0$
    \item Step 4: maximize $\lambda_n$, i.e.,  $\nabla_\lambda\mathcal{L}(x_n,y_n,\lambda) = 0$
    \item Step 5: Repeat steps 1-4 until the error is minimized
\end{itemize}

For our particular example we have that $x=p$, $y=A=\text{diag}(\varepsilon-1)$, and $f(p,A) = 1/2||(A^{-1}-G)p-e_\text{inc}||^2$ and $g(p) = 1/2||G_\text{pr}p-e_\text{obj}||^2$, and Lagrange function reads 
\begin{equation}
    \mathcal{L}(p,A,\lambda) = ||(A^{-1}- G)p-Ae_\text{inc}||^2 +\lambda||G_p p - e_{obj}||^2
\end{equation}
The corresponding algorithmic steps are: 
\begin{itemize}
    \item Step 0: initial $p_0$ and $\lambda_0$
    \item Step 1: minimize $A_n$ via $\nabla_A\mathcal{L}(p_{n-1},A,\lambda_{n-1}) = 0$ 
    \begin{itemize}
        \item We have that $\nabla_A\mathcal{L}(p_{n-1},A,\lambda_{n-1}) = \nabla f(p,A)^{*}||(A^{-1}-G)p-e_\text{inc}||$ ($*$ is complex conjugate). This expression lead to $A_n = p/(Gp-e_\text{inc})$. In practice this is a simple calculation since $A$ is a diagonal matrix, i.e., $A=\text{diag}(\varepsilon-1)$.
    \end{itemize}
    \item Step 2: project $A_n$ into $A_n \in \mathbb{R}$ and $0 \leq A \leq 4$ (for the range $\varepsilon\in[1,5]$)
    \begin{itemize}
        \item this is the point where essentially the required properties and bound of the permittivity can be implemented. These bounds or constrains can be general
        \item the above projection is rather a simple projection that does not guarantee always the minimum within the projection domain. 
        A more accurate projection would be of the form $A_{n}=\text{proj}[A_{n-1}-\eta\nabla_A||(A_{n-1}^{-1}- G)p_{n-1}-e_\text{inc}||^2]$.
    \end{itemize}
    \item Step 3: minimize $p_{n}$, i.e.,  $\nabla_p\mathcal{L}(p,A_{n},\lambda_{n-1}) = 0$ (DCM metadevice).  
    \begin{itemize}
        \item $p_{n} = K_n^{-1} e_\text{L}^n$  
        \item $K_n=(A^{-1}_n-G)^*(A^{-1}_n-G)+\lambda_{n-1}G^*_\text{pr}G_\text{pr}$
        \item $e^L_{n}=\lambda_{n-1}G_\text{pr}^*e_\text{obj}+(A^{-1}_n-G)^*e_\text{inc}$
        \item The matrix inversion $p_{n}$ is performed with our DCM metadevice
        \item Due to noise error a simple weighted average filtering is applied, i.e., $p_{n}=(1-\alpha_F) p_{n-1}+\alpha_F p_{n}$ with $\alpha_F=0.25$
    \end{itemize} 
    \item Step 4: maximize $\lambda_n$, i.e.,  $\nabla_\lambda\mathcal{L}(A_n,p_n,\lambda) = 0$ 
    \begin{itemize}
        \item This maximization can be calculated by a simple gradient descent, i.e., $\lambda_n = \lambda_{n-1}+\eta\left(\nabla_\lambda\mathcal{L}(p_{n},A_{n},\lambda)-\delta\right)$ or $\lambda_n = \lambda_{n-1}+\eta\left(||G_p p_n - e_{obj}||^2-\delta\right)$
        \item Notice that this is an gradient \emph{ascent} since we assume $\eta>0$, therefore we maximize the problem. 
    \end{itemize}
    \item Step 5: Repeat steps 1-4 until the error is minimized. In our case we used the following error
    \begin{itemize}
        \item $||e_\text{sca}-e_\text{obj}||^2/||e_\text{obj}||^2$  
    \end{itemize}
\end{itemize}
Note that the quantities $\eta$ and $\delta$ are the step and minimal error quantities that are user determined. The whole process stop either when $\lambda$ reaches a plateau, or when the required error criterion is met. The optimization goal was set as $\frac{||e_\text{sca}-e_\text{obj}||^2}{||e_\text{obj}||^2}<\delta$, where $e_\text{sca} = G_p p_\text{m}$ with $p_\text{m}=(A_m^{-1}- G)^{-1}e_\text{inc}$ being the final $m$-th evaluation of the iteration.

Notice that our approach has several similarities with the contrast source inversion method and other similar inverse scattering methods~\cite{Kleinman1992,Berg1997,Colton2013,Boutami2019}. 

Undoubtedly this approach is only one of the available methods for approximating the inverse design problem. This is rather an attempt to showcase the ability of our device for performing inverse design with desired objectives and constraints by exposing the crucial parts of the algorithm, such as the matrix inversions. This part is usually implicit within commercially available FDTD or FEM software. Hence here we developed our own methodology so we can have deeper inspection to quantities. As a remark, the field of inverse design and inverse scattering is a very rich field with a plethora of methodologies that try to address similar problems~\cite{Li2022}. 

\section{RF Design, PCB, Device Implementation}
\begin{figure}[h] 
        \centering
        \includegraphics[width=1\textwidth]{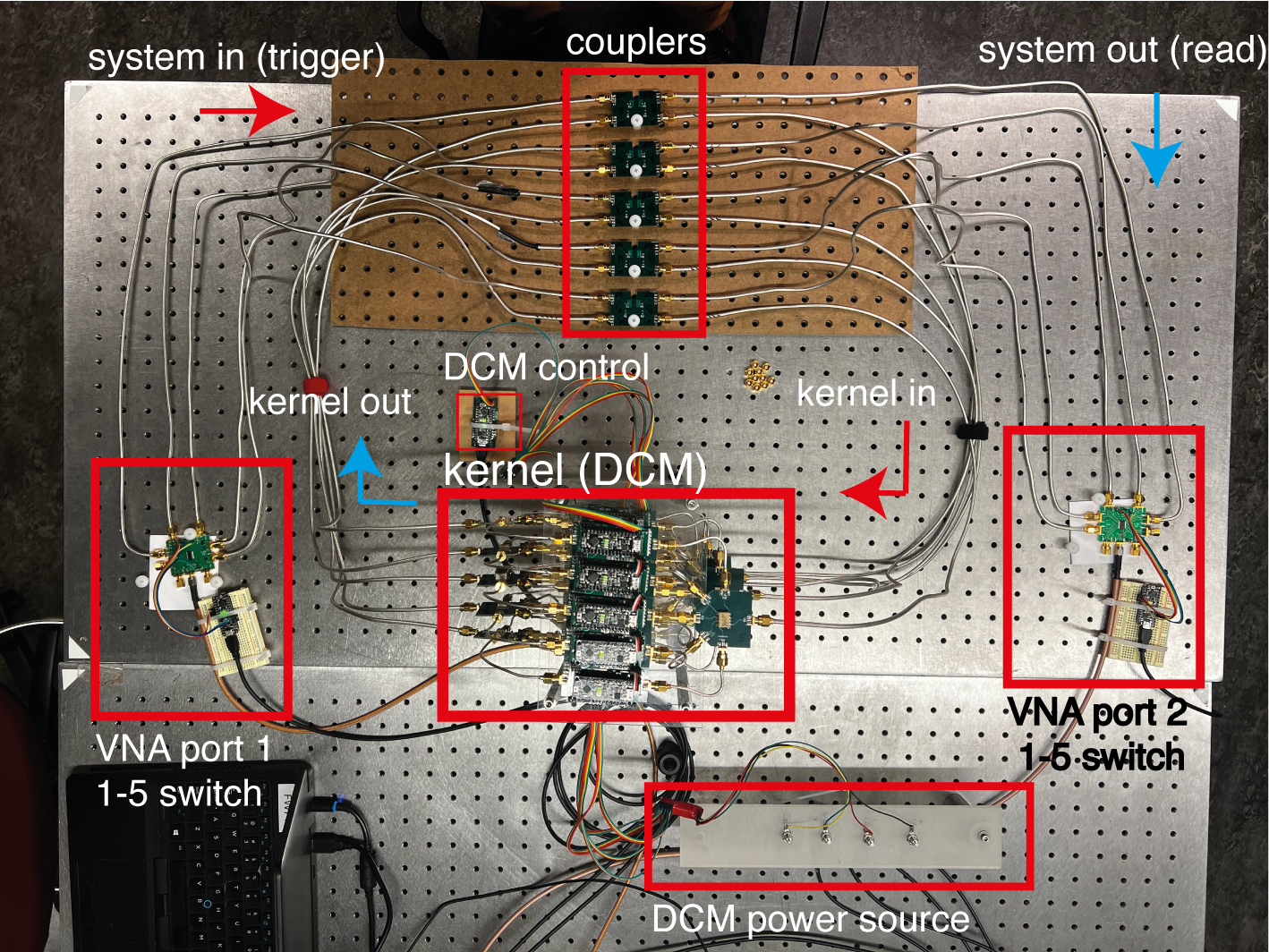}
        \caption{Photograph of the experimental setup with the corresponding components.}
        \label{fig:all}
    \end{figure}

A photograph of the experimental setup is shown in Fig~\ref{fig:all}, where all parts are designated accordingly.  
\subsection{Measurement}
Measurements were performed using an ENA-5071C two port VNA.  
In order to avoid the saturation of the amplifiers (multiplier module) the VNA power level was set to be $-20$dBm for the open loop configuration and $-10$dBm for the closed loop configuration.
The VNA was set to have an IF bandwidth of 10 kHz. 
The single frequency measurements (1601 point) at 45MHz with averaging applied after obtaining the measured signal from VNA. 

\subsection{Multiplier}\label{Multiplier}
	
	\begin{figure}[h] 
        \centering
        \includegraphics[width=0.5\textwidth]{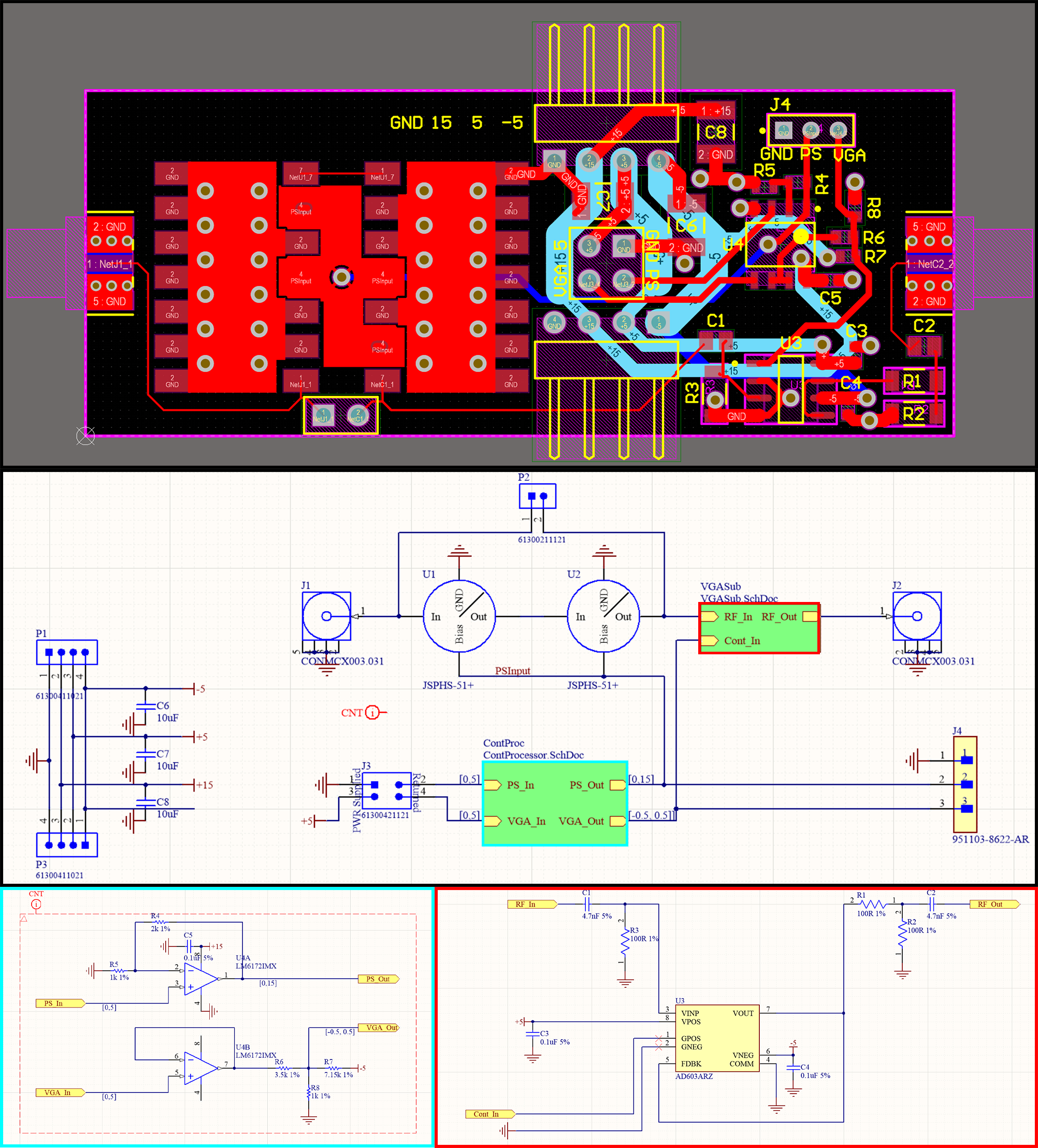}
        \includegraphics[width=0.5\textwidth]{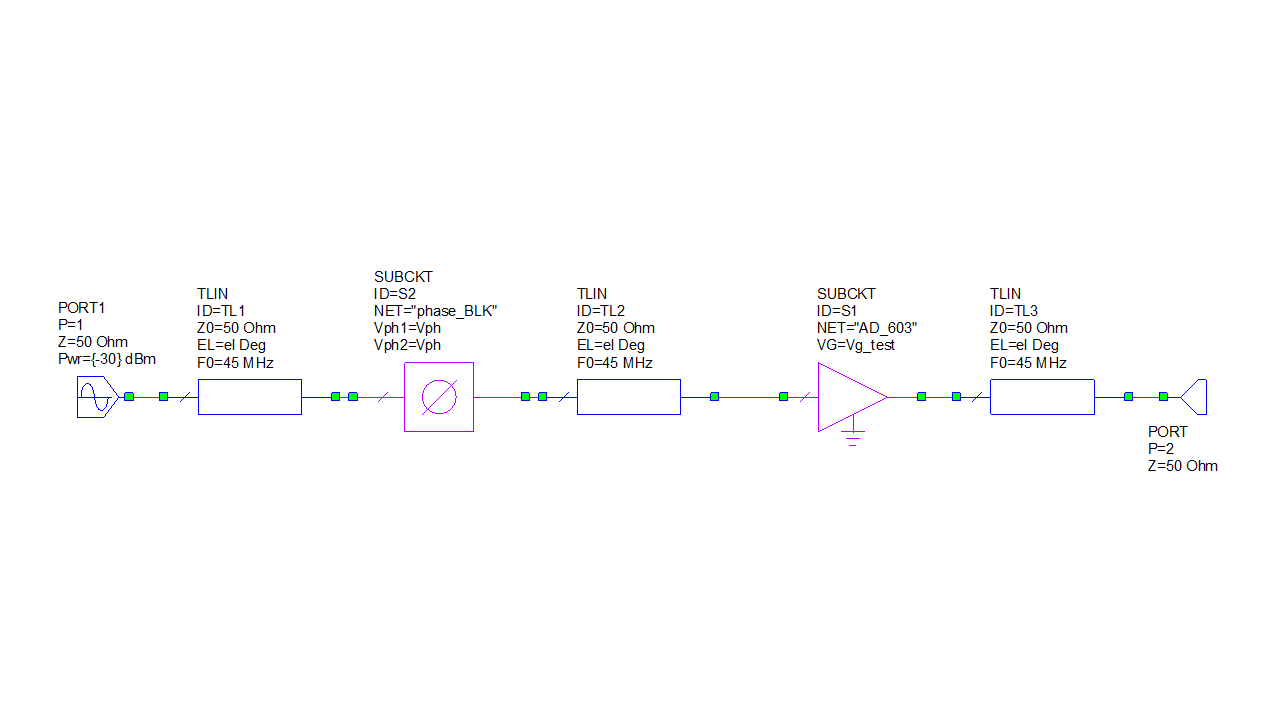}
        \caption{Schematics for the multiplier: The PCB layout design (top figure), and the corresponding subcircuit (pictures from Altium$^{\text{\textregistered}}$). Bottom figure represent the AWR Microwave Office$^{\text{\textregistered}}$
 schematic with the realistic data}
        \label{fig:mult}
    \end{figure}
    
	The schematic of the multiplier is depicted in Fig.~\ref{fig:mult}. 
	The multiplier was designed to perform multiplication on the incoming complex amplitude such that a new complex amplitude is rendered at the output.
	In other words, the output is $V_\mathrm{out} = z V_\mathrm{in}$.
	This involves changing both the amplitude and phase of the incoming signal.
	Phase change was performed using a pair of serially connected Minicircuit JSPHS-51+ Phase Shifters (PS).
	Each phase shifter provides slightly over 180 degrees of rotation.  
	The amplitude change was performed using the Analog Devices AD603ARZ Variable Gain Amplifier (VGA).
	The Multiplier design contain the appropriate loads such that both the input and output of the device externally appears as 50 Ohm.
	
	Both of these devices are controlled using analog voltages with ranges of $[-0.5\mathrm{V}, +0.5\mathrm{V}]$ and $[0\mathrm{V}, 12\mathrm{V}]$ for the VGA and PS, respectively.
	In order to create a common control mechanism, op-amp level shifting circuits were used to put these on a common $[0\mathrm{V}, 5\mathrm{V}]$ interface.
	The Multiplier board has a connection that allows for a daughter board.
	The daughter board is supplied with 0V and +5V and is responsible for returning two control voltages in the range of $[0\mathrm{V}, 5\mathrm{V}]$.  
	
	This simple interface allows for a number of possible control schematics.  
	At its most simple scenario, the control board can consist of a pair of potentiometers.  
	However, we will present another control board which utilizes a microcontroller to receive UART input and render the two analog voltages.
	
	The VGA's dynamic range could be shifted using an external resistor.  
	This was set so that the Multipliers's range (including load elements, PS losses, etc) was [-30dB -- +17dB].  
	The multiplier effectively saturates if the input is greater than -10dBm.
	Therefore for all measurements the reference input signal that was used was -30dBm for avoiding any saturation effects.
	
	It should be stated that the VGA imparts a varying phase change and the PS pair imparts an amplitude change. This will be addressed later.

\subsection{1-5 splitter (5-1 combiner)} 
    
	\begin{figure}[h] 
        \centering
        \includegraphics[width=0.5\textwidth]{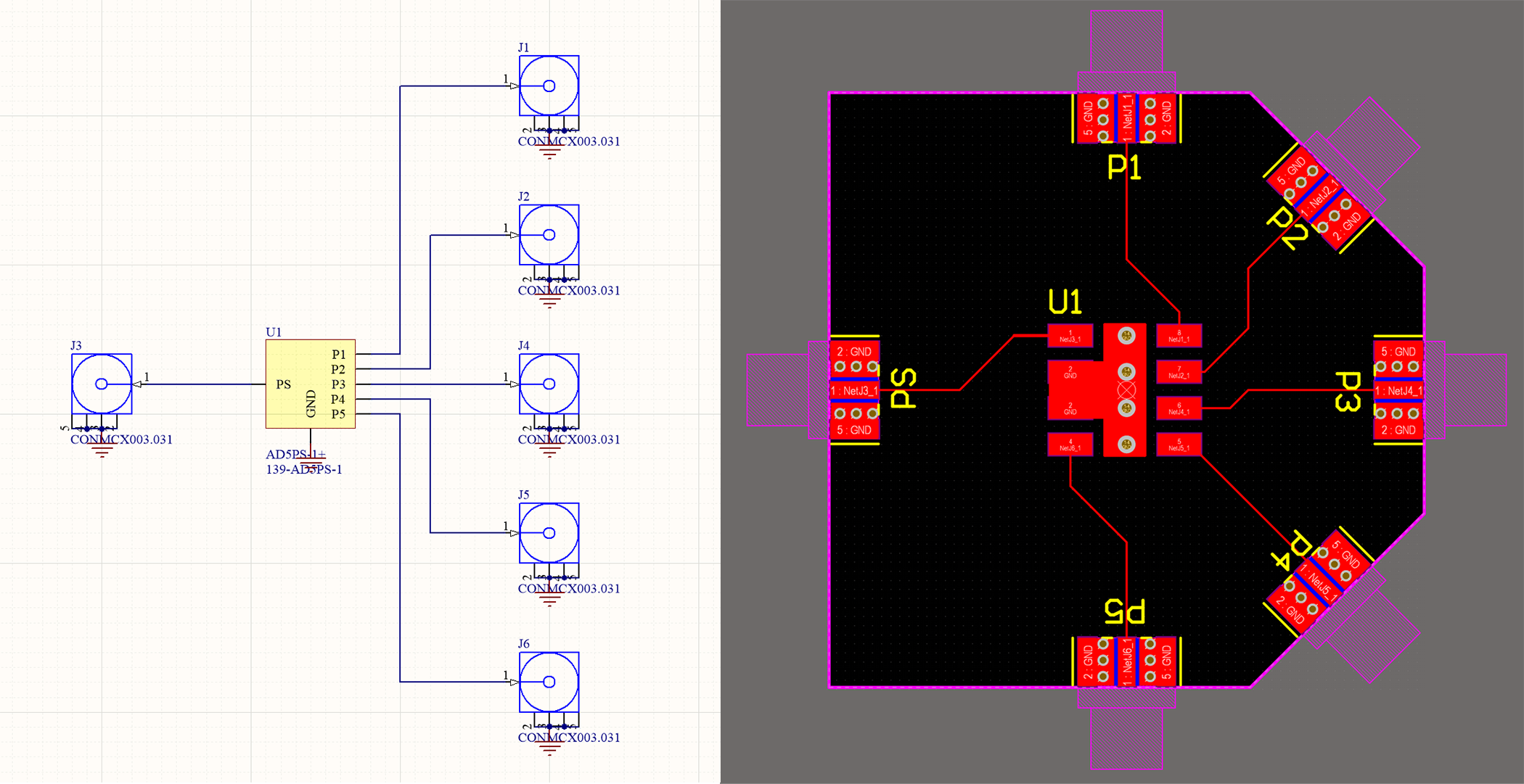}
        \caption{Schematic and layout for 1-5 splitter based on the Minicircuits AD5PS-1+  (pictures from Altium$^{\text{\textregistered}}$)}
        \label{fig:5-1}
    \end{figure}
    
    The schematic of the 1-5 splitter is depicted in Fig.~\ref{fig:5-1}. 
    An ideal passive $n$-way splitter is comprised of a summation port and $n$ feed ports.  
    The scattering parameters are expected to be reciprocal such that for the $i^\mathrm{th}$ feed port $|S_{\mathrm{S}i}|^2 = |S_{i\mathrm{S}}|^2 = 1/n$ and all other elements within the matrix are zero. 
    Due to losses, a real splitter will fall short of this precise definition.  
    Our splitter was based on the Minicircuits AD5PS-1+, which yielded good performance at 45MHz with approximately -7.2dB split ratio for all outputs.Note that $1/5 \approx -7.0dB$.

\subsection{Feedback coupler}
    
    \begin{figure}[h] 
        \centering
        \includegraphics[width=0.5\textwidth]{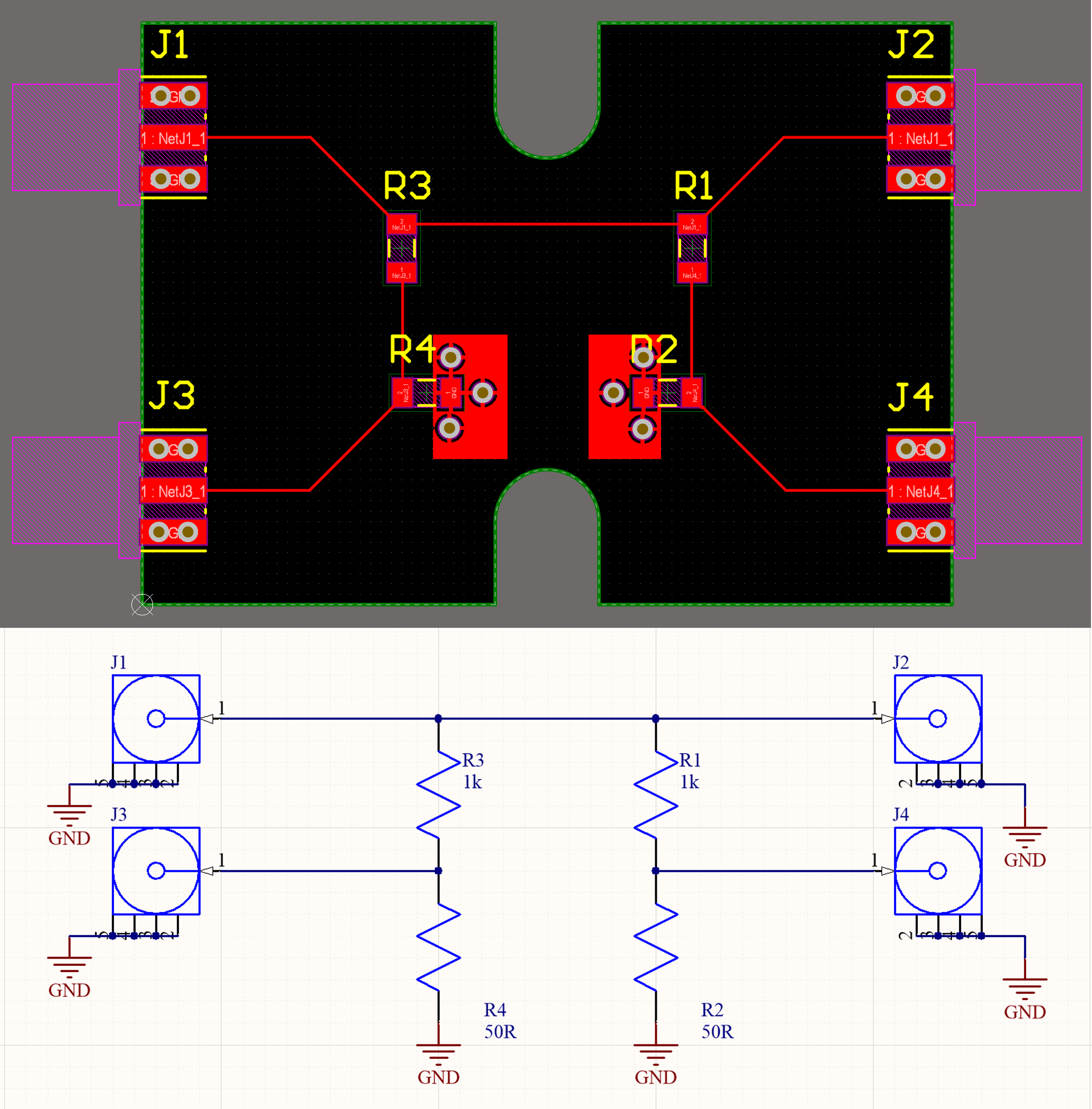}
        \caption{Schematic and layout for the feedback coupler (pictures from Altium$^{\text{\textregistered}}$)).}
        \label{fig:FC}
    \end{figure}

    The schematic of the feedback coupler is depicted in Fig.~\ref{fig:FC}
    The Feedback coupler must perform several tasks.
    \begin{itemize}
      \item Provide near unity feedback
      \item Introduce the input signal
      \item Sample the output signal
    \end{itemize}
    Therefore, the feedback coupler is a four-port device wherein the primary path has near unity transmission such that the feedback is strong.
    
    \subsection{Switches}
   
    In order to replicate having a 10 port VNA, we utilized two demo boards (EV1HMC253AQS24), which acted as RF SP8T RF switches, i.e., an analog multiplexer.
    For one SP8T, we utilized five of these ports for illuminating the bank of five couplers.
    The other SP8T was used to receive signals from the couplers.
    The remaining three ports on each were used for system sanity checks.
    Note that the stock high-pass 100pF capacitors on these boards were switched to 470pF for better transmission at 45MHz.
    
    While the off ports were nominally matched to 50 Ohm from DC-2.5GHz, there was significant reflections.  
    Deeper inspection of the datasheet indicated that the "off" ports were only matched above 500MHz.
    Measurements indicated that the off ports were approximately "open" at the design frequency and therefore reflections from the off ports could be significantly reduced with parallel 50Ohm terminations.
    However, this was not done as this it would have reduced power within the system on the "on" port.
    Rather, we note that any polluting signal from these "open" off ports will have crossed through the coupler twice.
    Due to the small coupling coefficient of the feedback coupler, these values will have become very small.

    The VNA was calibrated to the end of the switch ports.
    Measurements indicated that transmission through each of the switch ports was similar enough as to not warrant individual calibrations on each.

    Each of the switches was actuated by three digital inputs to address the $2^3 = 8$ ports on each switch.  
    These digital signals were created by a micro-controller which was programmed to respond to UART commands from an attached computer.  
    Code is available at \path{github.com/brianedw/RFMath/Arduino/mcu_control_V2/mcu_control_V2.ino}.

    \subsection{Micro Controller Unit (MCU)}    
    The two analog input control voltages for each Multiplier was created by an MCU Control Board, which attached directly to the Multiplier.  
    The heart of this board is a Metro-Mini MCU.  
    
    Each control line was connected to both an 8-bit PWM DAC pin (labeled ``fast'') and a 10-bit PWM DAC pin (labeled ``slow'').  
    While both pins connected to the control line through a high-pass filter, the fast DAC utilized a lower capacitance and resistance than that of the slow DAC.  
    During a set operation, both pins would drive to their appropriate values, during this time, the behavior of the collective output would be dominated by the fast DAC and rapidly converge, but exhibit large ripples.
    After 20ms, the fast PWM DAC pin would switch to a high-impedance state, leaving the voltage to settle in the remaining difference utilizing the slow DAC alone.
    The high-pass filter was designed to maintain accuracy of 10bit.
    Since the Metro-Mini is a 5V compliant device, the generated voltages nicely matched to the expected inputs of the Multiplier.  
    
    Each MCU board had two 3-pin UART input connectors.  
    These were shorted such that one could be used to receive a command from "upstream" while the other would effectively passively repeat the signal.  
    Additionally, each MCU board had two 3-pin UART output connector which were similarly shorted together, allowing it to transmit the same message to two devices.  
    Each MCU was programmed with a unique identification number. 
    Upon receiving a UART command, it would either act on that command or repeat the command on its output UART pins for downstream devices.  
    This input/output configuration created a lot of possibilities for control topologies.  
    However, in practice we found that we could use a single MCU board (no multiplier attached), as a bridge between the computer and the array of MCU Boards and that this array could all be connected in parallel such that the output of the bridge was effectively driving 25 inputs.
    Note that the required time complexity is of the order of $\mathcal{O}(n^2)$. 
    Possibly this complexity can be further reduced by implementing different connectivity schemes than the simple serial one that we used. 
    Code is available at \path{github.com/brianedw/RFMath/Arduino/mcu_control_V2/mcu_control_V2.ino}.

\section{Tuning/Calibration}

As stated in \ref{Multiplier}, the VGA has a minor effect on the phase and the PS has a minor effect on the amplitude.  
In other words, the phase and amplitude responses are coupled.  
Additionally, other systematic errors are present such as nonidealities in the level shifting circuits due to resistor tolerances.  
When connected in a network that includes RF jumper cables of varying length, there will also be phase shifts that naturally arise.  
In short, the relationship between the control voltages and the response of the Multiplier \emph{in situ}, are repeatable, but difficult to predict without developing a more complex model.

We found that an effective strategy to capture, model, and invert the relationship between control voltages and system response goes as follows.
\begin{enumerate}
    \item A collection of Multipliers are swept across their input values to map the relationship between control voltage and complex multiplier response.
    \item These responses were analyzed using Principle Component Analysis (PCA)~\cite{jolliffe2002principal}.
    \item The multipliers were assembled into the open-loop configuration and the response of the entire open-loop network was measured under many sets of input control voltages.
    \item These results were compared to a theoretical model of the network wherein the weights of the components could be adjusted until the theoretical results matched the experimental results.
    \item With accurate PCA weights in hand, the Multipliers can be immediately adjusted to achieve a desired multiplication factor by inverting the model to achieve any open-loop kernel.
    \item Additional refinement can be obtained by changing the device configuration into the closed loop, which now includes the feedback couplers.
          Again, we measure the response of the closed-loop network under many input conditions.
    \item We further refine the PCA weights of each multiplier to match this more demanding data set.  This becomes our final device model for both the open- and closed-loop configurations.
\end{enumerate}

We will go into detail on each one of these items in the following sections.

    \subsection{Multiplier PCA}

    A collection of 35 multipliers were each mapped using the MCU control boards, capable of 10-bit resolution on both control voltages.  
    The mapping occurred with a grid of values based on [0, 11, ..., 1012,  1023] on both controls.
    Ideally, the mapping of two Multipliers would yield identical responses.  
    However, for all the reasons stated above, they do not.  
    All of the mappings were compared using a complex domain PCA analysis.  
    Typically, in PCA, one would examine deviations from the mean, but here we take another approach.
    Rather, the collection of mappings were analyzed directly to yield a set of 4 PCA components.
    The response of any individual Multiplier could then be found as the linear superposition of these components given by:
        \[ m(d_\mathrm{vga}, d_\mathrm{PS}) = \sum_{i=0}^{3} w_i c_i(d_\mathrm{vga}, d_\mathrm{PS}) \]
    The term $c_0(d_\mathrm{vga}, d_\mathrm{PS})$ is effectively the ``average'' response scaled by a complex factor, while the next several components represent likely deviations due to the systematic errors described above.  
    Within a PCA analysis the final PCA components (i.e. $c_{34}(d_\mathrm{vga}, d_\mathrm{PS})$, not shown) should be nearly pure noise.  
    We found that only the first four terms were needed to effectively model any given Multiplier.
    
    Given any randomly chosen multiplier, we can find the complex valued PCA weights $w_i$ through a least-squares analysis.  
    As opposed to the "deviation from the mean" approach, the above formulation is particularly useful for RF engineering.  
    While the Multipliers were measured directly at their input and output ports and analyzed as such, the model can easily account for the addition of RF cables which would provide attenuation and phase rotation.  
    These will appear as a complex scaling of all of the components weights and the Multiplier's behavior (RF jumpers cables included) can still be captured as the simple linear superposition of the PCA components.  
    In fact, any losses or phase rotations along the Multipliers flow path can be incorporated into these weights.  
    Therefore, we do \emph{not} characterize the individual multipliers, but delay this until the architecture is fully assembled, as described in the next section.
    
    Regardless, we will use least-squares to find the set of $w_i$ which characterizes the average multiplier response. 
    We call these the ``base weights''.
    
    \begin{figure}[h] 
    \centering
    \includegraphics[width=1.0\textwidth]{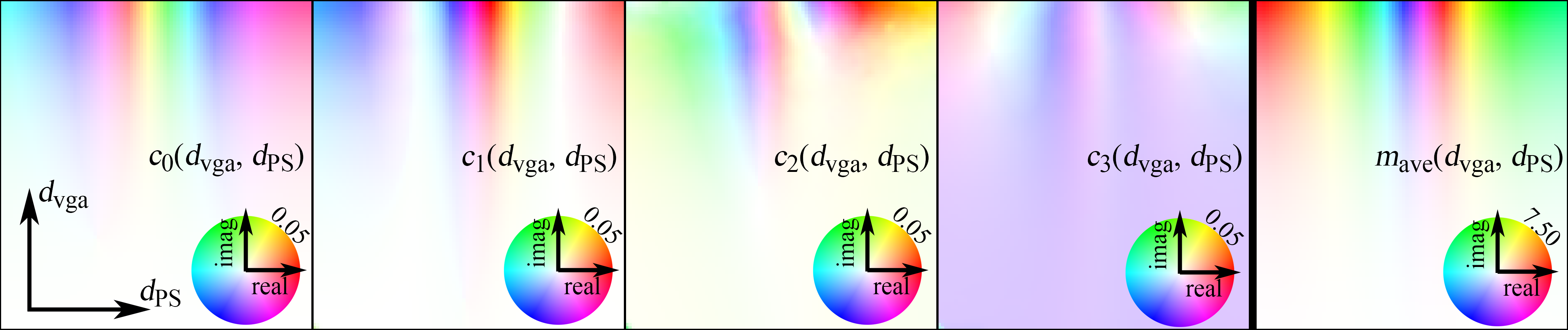}
    \caption{PCA Components and Average Multiplier Response.  The first four panels represent $c_0(d_\mathrm{vga}, d_\mathrm{PS})$, $c_1(d_\mathrm{vga}, d_\mathrm{PS})$, $c_2(d_\mathrm{vga}, d_\mathrm{PS})$, and $c_3(d_\mathrm{vga}, d_\mathrm{PS})$ and have a maximum saturation of 0.05.  The final image shows the response of the ``average'' Multiplier with a maximum saturation at 7.5}
    \label{fig:my_label}
    \end{figure}
    
    \subsection{Open-Loop Device Fitting}
    
    The goal of the this section is to determine the PCA weights that characterize each Multiplier \emph{in situ}, so that the system errors can be captured and modeled. 
    The open-loop DCM system was fully assembled including jumper cables, splitters, and couplers.  
    All multipliers within the array were set to the same input value $(d_\mathrm{vga}, d_\mathrm{PS})$ pair.  
    The transmission matrix of the system was then measured.  This was repeated for all possible combinations of 10 evenly spaced values in the range $[0, 1023]$ to yield 100 measured transmission matrices, $\mathcal{T}_\mathrm{meas}$.
    Note that not all of these 100 transmission matrices represent "passive" operators.
    
    The same system was modeled using Scikit-RF, wherein the following assumptions were made:
    \begin{itemize}
      \item The 5-1 splitters were ideal such that power was evenly split with no phase
      \item All jumpers were zero-length
      \item The coupler feedback path was ideal with no power removed
      \item The multipliers were all assumed to be ``average'' and the their responses were assumed to be given by the ``base weights''.
      \
    \end{itemize}
    The system was simulated using SciKit-RF for each input pair $(d_\mathrm{vga}, d_\mathrm{PS})$ to yield 100 measured transmission matrices $\mathcal{T}_\mathrm{sim}(\overline{w})$, which are naturally a function of each Multipliers PCA weight. 
    We can then define an error $\mathrm{error}(\overline{w}) = \left| \mathcal{T}_\mathrm{sim}(\overline{w}) - \mathcal{T}_\mathrm{meas}\right|^2$ and optimize $\overline{w}$ until that error is minimized.  
    It should be noted, that with only four PCA weights per Multiplier, in theory, only 4 transmission matrices are required to fully define the system.  
    Using 100 helps guarantee that normal measurement noise does not unduly influence the fitting.  
    Additionally, if a low error can be achieved across 100 measurements using only 4 weights, then we can be confident that the model was sufficient to capture the entire open loop system response, $\mathbb{K}$.
    
    \subsection{Setting the Open Loop System Response}
    
    Given a desired open-loop system response, $\mathbb{K}$, we need to calculate the necessary multiplier values for the DCM architecture, $m_{i,j}$, gathered to form $\mathbb{M}$.  
    In this case, the simplicity of the DCM architecture makes this trivial.  
    If we assume an idealized passive five port splitters such that given an input of 1W at the summation port, s, we will observe 1/5W on each branch port, $i$.  
    Put in terms of Scattering Parameters, $S_{\mathrm{s},i} = 1/\sqrt{5}$ and via reciprocity $S_{i, \mathrm{s}} = 1/\sqrt{5}$.  
    Since we have such splitters at the input and output of the Multiplier array, $\mathbb{K} = (1/\sqrt{5}) \mathbb{M} (1/\sqrt{5})$ and therefore $\mathbb{M} = 5 \mathbb{K}$.  
    Note that since we fitted the PCA weights of the Multipliers under the assumption of ideal components, it is appropriate to assume ideal components here.
    
    With each of the desired $m_{i,j}$ in hand to achieve a given $\mathbb{K}$, the next step is to determine the required $(d_\mathrm{vga}, d_\mathrm{PS})$.  
    This can be done using a number of function inversion schemes such a gradient descent.  
    In practice, this could be very fast as it is likely that in many applications, each new $\mathbb{M}$ will be a small step from the previous $\mathbb{M}$ and therefore each multiplier will change only slightly.
    
    \subsection{Closed-Loop Device Fitting}
    
    Due to the recursive nature of the closed-loop configuration (Matrix Inversion), the accuracy requirements are more stringent than for the open-loop configuration.  
    Moreover, additional degrees of freedom are introduced in the form of coupler coefficients.
    These can be considered part of $\overline{w}$.
    In short, the devices must be fitted again.
    
    We will employ a similar strategy as was used in the \emph{Open-Loop Device Fitting}.
    Using the open-loop calibrated device models, a sequence of randomly generated passive transmission matrices, $\mathbb{K}$, are shown to the system.
    Note, unlike the open-loop matrices, in order to guarantee convergence, these matrices must be passive.
    We model the closed-loop system using Scikit-RF.
    Using the open-loop weights as a starting point, we optimize the multiplier weights and coupling coefficients until the simulated $\mathcal{T}_\mathrm{sim}(\overline{w})$ matches the measured $\mathcal{T}_\mathrm{meas}$.
    This represents a small, but necessary, refinement from the open-loop device model and can be used for both open- and closed-loop applications.
    
    \subsection{Setting the Closed Loop System Response}
    
    Setting the closed-loop system response, $\mathbb{K}$ is identical to setting the open-loop response.
    In both cases, each desired $m_{i,j}$ is used to find the required $(d_\mathrm{vga}, d_\mathrm{PS})$ using a function inversion scheme.
    
\section{System Accuracy}
\begin{figure}[h] 
    \centering
    \includegraphics[width=0.65\textwidth]{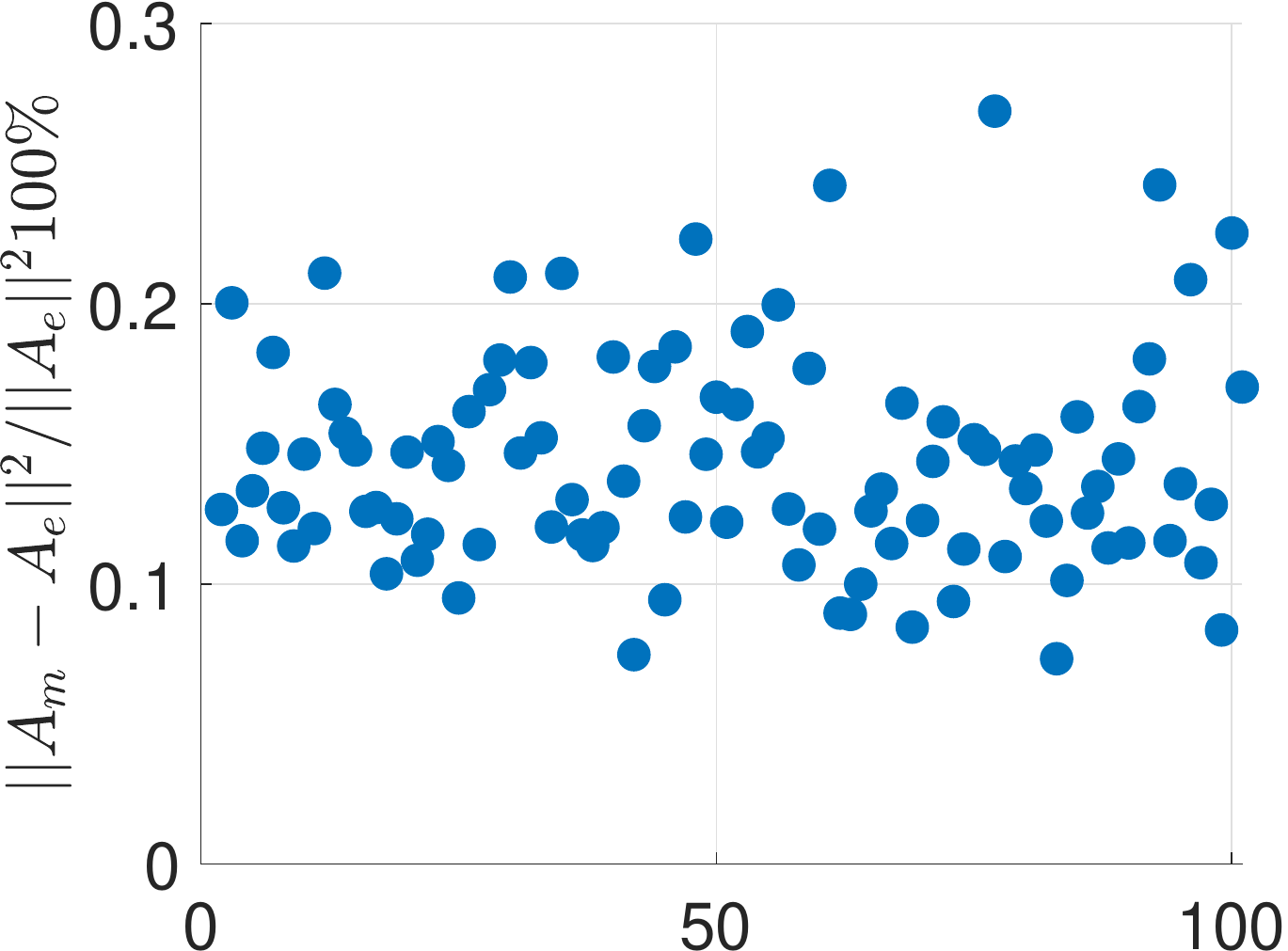}
    \caption{The error between the exact and the measured matrices, open loop configuration, for 100 random complex matrices.}
    \label{fig:big_kappa1}
    \end{figure}

\begin{figure}[h] 
    \centering
    \includegraphics[width=0.65\textwidth]{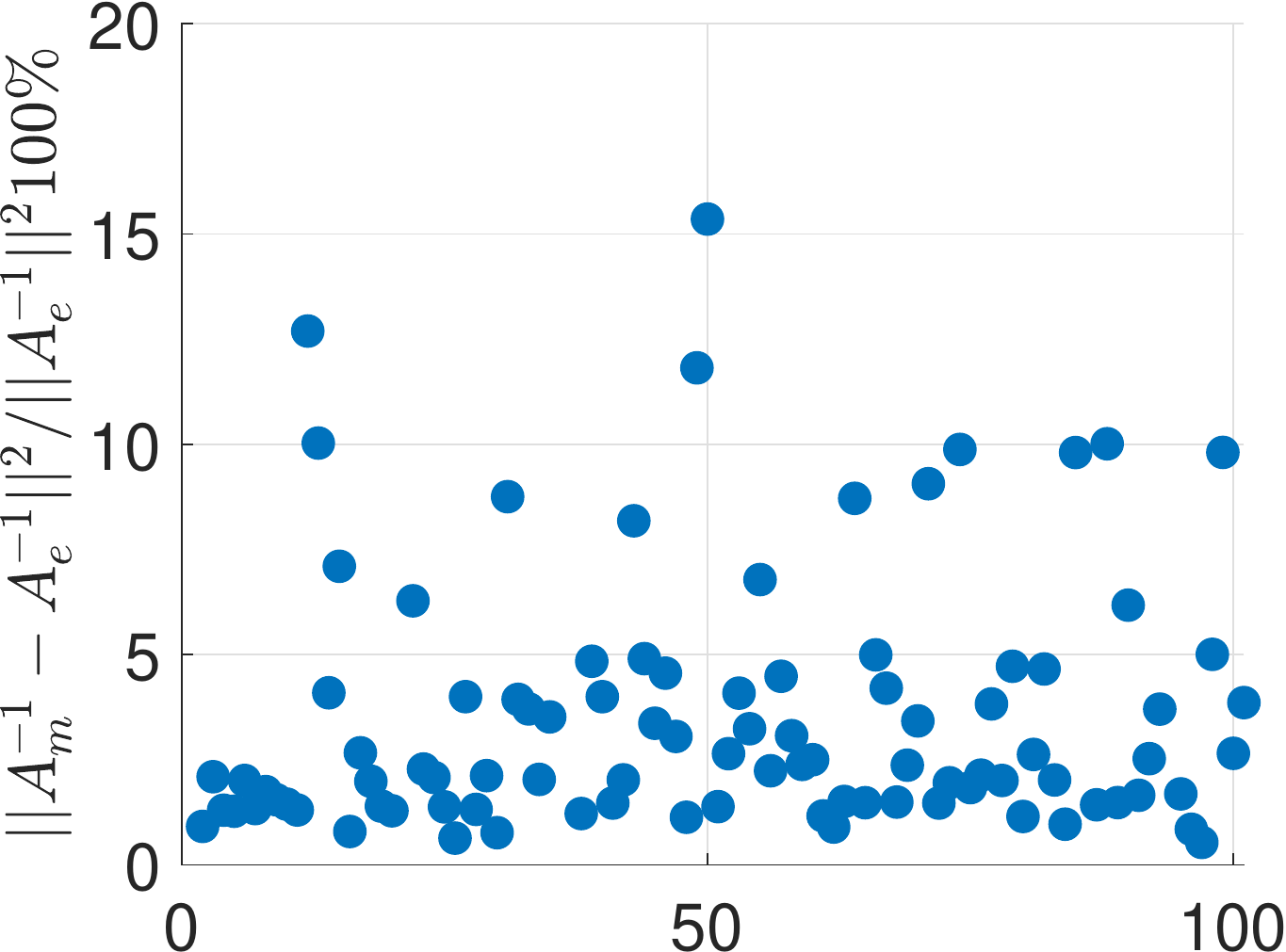}
    \caption{The error between the exact and the measured matrices, closed loop configuration (matrix inversion), for 100 random complex matrices.}
    \label{fig:big_kappa2}
    \end{figure}

\begin{figure}[h] 
    \centering
    \includegraphics[width=0.65\textwidth]{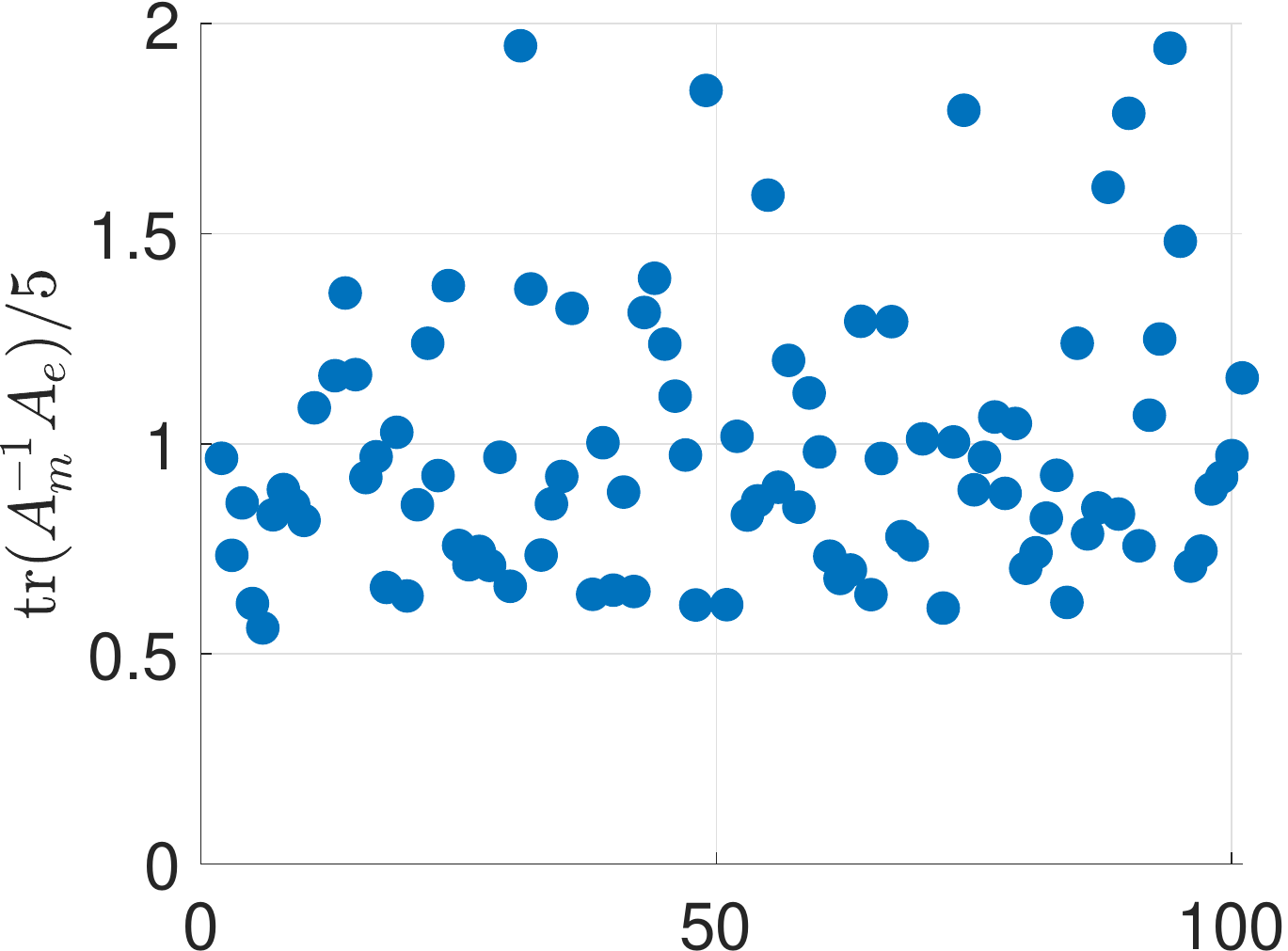}
    \caption{The fidelity of the matrix inversion expressed in terms of the normalized trace of the $A_m^{-1}A_e$ product for 100 random complex matrices.}
    \label{fig:big_kappa3}
    \end{figure}

We performed an open loop measurement on 100 complex-valued random matrices with (eigenvalues) values within the unit circle. For these, we configured the open-loop with the target (or ideal kernel) $A_e$ and retrieved the measured results $A_m$. We define as error the quantity 
\begin{equation}
\frac{||A_m-A_e||^2}{||A_e||^2}100\%
\end{equation}
In Fig.~\ref{fig:big_kappa1}, we can see the difference between the two matrices for 100 random cases. We observe that all the results are within a $0.05-0.3\%$ percent error. Similarly, we performed the same error analysis for the same 100 random matrices, only this time on a closed-loop setup (matrix-inversion). The results (Fig.~\ref{fig:big_kappa2}) reveal that the error can climb up to 20\%, but for most of the results, we get a matrix inversion with less than 2\% error. Finally, we assess the matrix inversion fidelity by evaluating the trace of the $A_m^{-1}A_e$ product. Ideally the trace of the product $\text{tr}(A_m^{-1}A_e)/5 = 1$. In Fig.~\ref{fig:big_kappa3}, we observe that this product spans between $0.5-1.5$. However, for the particular examples we used in the manuscript, this accuracy can be maintained at reasonably high levels once error-correcting and filtering techniques are applied. Note that for the closed-loop case, the level of the measured voltage is in the order of $\mu$V, very close to the noise floor of the VNA device we used. For the open loop operation, the measured voltage was hundreds of mV. 

\section{System Transient Analysis}
    \subsection{Single Multiplier}
\begin{figure}[h] 
    \centering
    \includegraphics[width=0.65\textwidth]{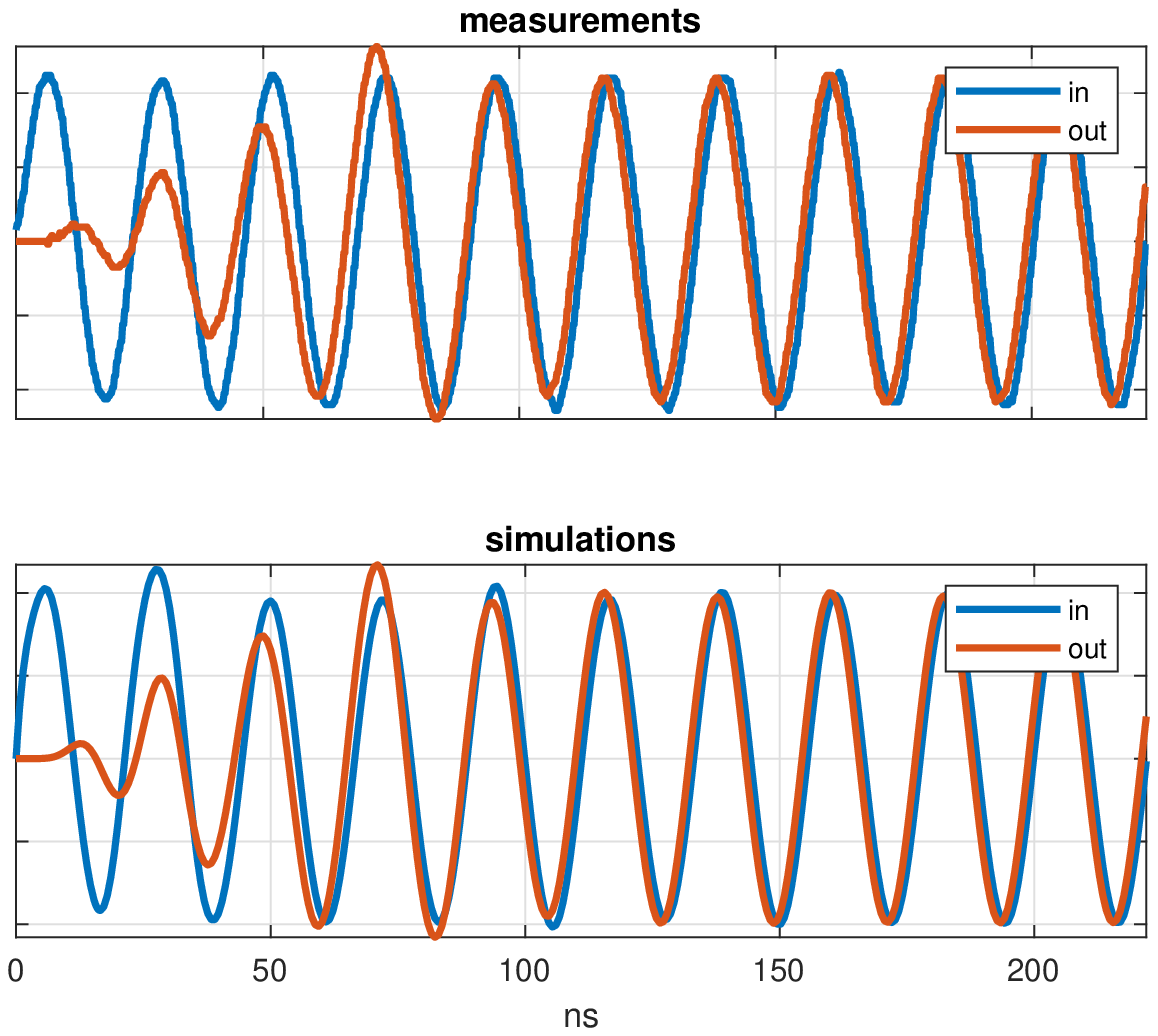}
    \caption{The transient response of a single multiplier module. The blue curves correspond to the input signal, while the red curves are the measured (top) and simulated (bottom) using AWR Microwave Office$^\text{\textregistered}$ results. The agreement is excellent. It is evident that it takes approximately 3 to 4 signal periods for the multiplier to obtain the desired output signal. Here we assumed small signal amplification (VGA voltage is +.05) and the phase shift voltage is 0V.}
    \label{fig:trans}
    \end{figure}

    In terms of the time response of the multiplier module the transient analysis reveal (Fig.~\ref{fig:trans}) that the module obtained the desired value approximately within 3-4 signal periods, i.e., $T=22.2$ns. The measurements were performed using the RIGOL DG4062 pulse generator (15 sinusoidal pulses at 45MHz), and the measured response extracted with the RIGOL DS1104 oscilloscope. 
    
    The open loop response is therefore assumed to be very close to the single multipliers response since both splitters and connecting cables introduce a small phase shift to the signal. The closed loop transient response is affected by both the multiplier timing and the condition number of the input matrix (kernel) as shown in~\cite{Tzarouchis2022}. 

    \clearpage 
\section{De-embedding the solution}

    \subsection{Open Loop}
    
        Let us define the open-loop response as
        \[
        V_\mathrm{out} = \mathbb{K} V_\mathrm{in}
        \]
        Note that this includes not only the DCM architecture (multipliers, splitters, jumpers), but also the through channel of the input/output coupler.
        In other words, the open-loop is defined using all of the components of the closed-loop.
        However, the loop has been broken "open" just after the coupler array and measured at this point.
        Since in the closed configuration, these measurement planes were coincident, upon "closing" the loop, these measurement will then represent the complete response of the loop.
        While a minor perturbation to the results, this definition assumes that the weakly coupled additional ports on the coupler are properly terminated.
        
        Let us further define response of only the DCM architecture as $\mathbb{K}'$.  When the system is in a closed loop configuration, this relates the vector exiting the coupler array ($V_4$) to the vector incident on the coupler array ($V_2$).
        \[
        V_2 = \mathbb{K}' V_4
        \]
        The coupler array introduces a small loss as the input is introduced and the output is measured.  The near unity transmission is named $\alpha_1$.    It is clear then that $\mathbb{K} = \alpha_1 \mathbb{K}'$.
        
    \subsection{Closed Loop}
        The closed loop response is fully defined by the open-loop response and the definition of the scattering parameters of the coupler.
        
        \begin{align} 
        V_2 &= \mathbb{K}' V_4 \label{DE1}\\
        V_3 &= \alpha_2 V_1 + \beta V_2 \label{DE2}\\
        V_4 &= \beta V_1 + \alpha_1 V_2 \label{DE3}
        \end{align}
        
        Our goal is to solve the equations for $V_4$, which represents the vectorial solution of the problem in question.  For the \emph{expected solution}, this should be done such that the solution depends only on the kernel $\mathbb{K}$ and the input vector ($V_1$).  For the \emph{measured solution}, this should be only in terms of the measured results ($V_3$) and the known input ($V_1$).
        
    \subsection{Expected Solution}
    
        We begin by applying the definitions above
        
        \begin{align*} 
        V_4 &= \beta V_1 + \alpha_1 V_2\\
        V_4 &= \beta V_1 + \alpha_1 \mathbb{K}' V_4\\
        V_4 &= \beta V_1 + \mathbb{K} V_4 \\
        \end{align*}
        
        and then solve the final equation for the $V_4$.
        \[
        V_4 = (I - \mathbb{K})^{-1} \beta V_1
        \]
    
    \subsection{Measured Solution}
        We begin with Eq \ref{DE2}:
        \[ V_3 = \alpha_2 V_1 + \beta V_2 \]
        and then solve it for $V_2$
        \[ V_2 = \frac{1}{\beta}V_3 - \frac{\alpha_2}{\beta}V_1 \]
        and then substitute the above into \ref{DE3}
        \[ V_4 = \beta V_1 + \alpha_1(\frac{1}{\beta} V_3 - \frac{\alpha_2}{\beta} V_1)\]
        and then simplify
        \[ V_4 = (\beta - \frac{\alpha_1 \alpha_2}{\beta}) V_1 + \frac{\alpha_1}{\beta} V_3\]
        Note that in many real world cases, the coupler will be defined such that we can assume $\alpha_2 \rightarrow 0$
        \[ V_4 = \beta V_1 + \frac{\alpha_1}{\beta} V_3 \]

\bibliography{scibib}
\bibliographystyle{Science}